\documentclass{ieeeaccess}
\usepackage{cite}
\usepackage{amsmath,amssymb,amsfonts}
\usepackage{algorithmic}
\usepackage{graphicx}
\usepackage{textcomp}
\usepackage[utf8]{inputenc}
\usepackage[T1]{fontenc}
\usepackage{grffile}
\usepackage{longtable}
\usepackage{wrapfig}
\usepackage{rotating}
\usepackage[normalem]{ulem}
\usepackage{makecell}
\usepackage{booktabs}
\usepackage{capt-of}
\usepackage{hyperref}
\usepackage{epigraph}
\usepackage{subcaption}
\usepackage{multirow,array}
\usepackage{rotating}
\usepackage{adjustbox}

\author{Joakim Skarding}
\date{\today}
\title{Foundations and modelling of dynamic networks using Dynamic Graph Neural Networks: A survey}
\author{\uppercase{Joakim Skarding}\authorrefmark{1}, 
\uppercase{Bogdan Gabrys\authorrefmark{1}, and Katarzyna Musial
}.\authorrefmark{1}}
\address[1]{Complex Adaptive Systems Lab, Data Science Institute, University of Technology Sydney, Sydney, NSW 2007, Australia}
\corresp{Corresponding author: Joakim Skarding (e-mail: joakim.skarding@uts.edu.au).}
\tfootnote{ This work was supported by the Australian Research Council, ‘‘Dynamics and Control of Complex Social Networks’’ under
Grant DP190101087.}

\begin{document}

\begin{abstract}
Dynamic networks are used in a wide range of fields, including social network analysis, recommender systems and epidemiology. Representing complex networks as structures changing over time allow network models to leverage not only structural but also temporal patterns. However, as dynamic network literature stems from diverse fields and makes use of inconsistent terminology, it is challenging to navigate. Meanwhile, graph neural networks (GNNs)  have gained  a  lot  of  attention  in  recent years  for  their  ability  to perform  well  on  a  range  of  network  science  tasks,  such  as  link  prediction  and  node  classification. Despite the popularity of graph neural networks and the proven benefits of dynamic network models, there has been little focus on graph neural networks for dynamic networks. To address the challenges resulting from the fact that this research crosses diverse fields as well as to survey dynamic graph neural networks, this work is split into two main parts. First, to address the ambiguity of the dynamic network terminology we establish a foundation of dynamic networks with consistent, detailed terminology and notation. Second, we present a comprehensive survey of dynamic graph neural network models using the proposed terminology.
\end{abstract}


\titlepgskip=-15pt
\maketitle

\section{Introduction}
\label{sec:org58993e1}
The bulk of network science literature focuses on static networks, yet every network existing in the real world changes over time. In fact, dynamic network structure has been frequently seen as a complication to be suppressed, to ease progress in the study of networks \cite{strogatzExploringComplexNetworks2001}. Since networks have been used as representations of complex systems in fields as diverse as biology and social science, advances in dynamic network analysis can have a large and far-reaching impact on any field using network analytics \cite{holmeTemporalNetworks2012}.

Dynamic networks add a new dimension to network modelling and prediction -- time. This new dimension radically influences network properties which enable a more powerful representation of network data which in turn increases predictive capabilities of methods using such data \cite{aggarwalEvolutionaryNetworkAnalysis2014,liRestrictedBoltzmannMachineBased2018}. In fact, dynamic networks are not mere generalizations of static networks, they exhibit different structural and algorithmic properties \cite{michailElementsTheoryDynamic2018}. 

This work is both broader and narrower in scope than previous works. The first part of this survey (section \ref{sec:dynnetrepresentation}) is broader in scope than related surveys and introduces dynamic networks and dynamic network models (referring to the 'foundations and modelling of dynamic networks' part of the title). The second part of this survey (section \ref{sec:deeplearningmodels} and section \ref{sec:dynamiclinkprediction}) is narrower in scope and more detailed than related surveys, and is a survey on dynamic graph neural networks (referring to the 'using Dynamic Graph Neural Networks' part of the title).

\textbf{Foundations of dynamic networks:}

Dynamic networks suffer from a known terminology problem \cite{holmeModernTemporalNetwork2015}. Complex networks which change over time have been referred to, among others, as; dynamic networks \cite{carleyInteroperableDynamicNetwork2007,zhouGraphNeuralNetworks2018}, temporal networks \cite{masudaGuideTemporalNetworks2016,holmeTemporalNetworks2012}, evolutionary networks \cite{aggarwalEvolutionaryNetworkAnalysis2014} or time-varying networks \cite{casteigtsTimeVaryingGraphsDynamic2011}. With models often working only on specific types of networks, a clear and more detailed terminology for dynamic networks is necessary. We describe dynamic networks foundations as well as propose and develop an associated taxonomy of dynamic networks to contextualize the models in this survey and enable a more thorough comparison between the models. We are unaware of any work with a comprehensive taxonomy of dynamic networks and therefore it can be considered as the first major contribution of this paper.

Dynamic networks is a vast and interdisciplinary field. Models of dynamic networks are designed by researchers from different disciplines and they usually use modelling methods from their fields. This survey provides a cross-disciplinary overview of dynamic network models. This overview is not intended to be seen as a dynamic models survey, but rather as a context for dynamic graph neural networks and as a reference point for further exploration of the field of dynamic networks modelling.

We consider a dynamic network to be a network where nodes and edges appear and/or disappear over time. Due to the terminology problem establishing a terminology and a clear definition of a dynamic network is a necessity for a survey of any kind of dynamic network models such as dynamic graph neural networks. In the process, we introduce a specific and comprehensive terminology that enable future works to forego the extensive definition process and simply apply our terminology.

Related surveys \cite{holmeTemporalNetworks2012,holmeModernTemporalNetwork2015,kazemiRelationalRepresentationLearning2019} focus either on specific kinds of dynamic networks, for example, temporal networks \cite{holmeTemporalNetworks2012,holmeModernTemporalNetwork2015} or on specific types of models, for example, representation learning \cite{kazemiRelationalRepresentationLearning2019,xie2020survey,barros2021survey}. We are unaware of any work which gives as complete a picture of dynamic networks and dynamic network models as we do. The first section is thus broader in scope than other surveys that focus on only one network type or one type of network model.

\textbf{Modelling dynamic networks using Dynamic Graph Neural Networks:}
A dynamic graph neural network (DGNN) is considered to be a neural network architecture that can encode a dynamic network and where the aggregation of neighbouring node features is part of the neural network architecture. DGNNs encode both structural and temporal patterns in dynamic networks. To encode structural patterns DGNNs often make use of a graph neural network (GNN) and for temporal patterns, they tend to use time series modules such as recurrent neural networks (RNN) or positional attention. Spatio-temporal networks (graphs where the topology is static and only node or edge features change \cite{jain2016structural}) are out of the scope of this survey and thus so are Spatio-temporal graph neural networks \cite{yu2017spatio,jain2016structural}.
 
DGNNs, like GNNs and other representation learning models, are versatile in which tasks they can be applied to. With different decoders and different data, different tasks are possible. In practice, so far DGNNs have been applied to similar tasks as GNNs, the most common of these tasks are node classification \cite{parejaEvolveGCNEvolvingGraph2019,sankarDynamicGraphRepresentation2018,xu2020inductive,rossi2020temporal} and link prediction \cite{chenGCLSTMGraphConvolution2018,parejaEvolveGCNEvolvingGraph2019,xu2020inductive,rossi2020temporal}, which both have diverse and interesting application across many disciplines. Link prediction may for example be applied in knowledge graph completion \cite{trivediKnowEvolveDeepTemporal2017,wu2020temp} or by recommender systems \cite{xu2020inductive,rossi2020temporal}. DGNNs have also been used for novel tasks such as predicting path-failure in dynamic graphs \cite{liPredictingPathFailure2019}, quantifying scientific impact \cite{zhou2020heterogeneous}, and detecting dominance, deception and nervousness \cite{wang2020generic}.

There are several surveys on graph neural networks \cite{zhouGraphNeuralNetworks2018,zhangDeepLearningGraphs2018,wuComprehensiveSurveyGraph2019} as well as surveys on network representation learning \cite{hamiltonRepresentationLearningGraphs2017, goyalGraphEmbeddingTechniques2018}, our work differs from theirs as we cover GNNs which encode dynamic networks. Kazemi \emph{et al.} \cite{kazemiRelationalRepresentationLearning2019}, Xie \emph{et al.} \cite{xie2020survey} and Barros \emph{et al.} \cite{barros2021survey} are the works most similar to this paper as they survey dynamic network representation learning. The distinction is that they survey the broader topic of representation learning on dynamic networks whereas we survey dynamic graph neural networks which is a subset of representation learning on dynamic networks. We thus survey a more narrow scope than dynamic representation learning surveys and a different network type from the GNN surveys which focus on static networks \cite{zhouGraphNeuralNetworks2018,zhangDeepLearningGraphs2018,wuComprehensiveSurveyGraph2019}. Wu \emph{et al.} \cite{wuComprehensiveSurveyGraph2019} and Zhou \emph{et al.} \cite{zhouGraphNeuralNetworks2018} also survey spatio-temporal graph neural networks, which encode spatio-temporal networks (static networks with dynamic node attributes).

This survey's contributions are: (i) A conceptual framework and a taxonomy for dynamic networks, (ii) an overview of dynamic network models, (iii) a survey of dynamic graph neural networks (iv) an overview of how dynamic graph neural networks are used for prediction of dynamic networks (dynamic link prediction).

This work follows the encoder-decoder framework used by Hamilton \emph{et al.} \cite{hamiltonRepresentationLearningGraphs2017} and is split into three distinct sections each building upon the previous one.
\begin{enumerate}
    \item Section \ref{sec:dynnetrepresentation} is a discussion on dynamic networks. It serves as a foundation to the following sections. In this section we explore different definitions of links and introduce a novel dynamic network taxonomy. We also give a brief overview of the dynamic network model landscape, which contextualizes the rest of the survey.
\item Section \ref{sec:deeplearningmodels} is a survey of the deep learning models for encoding dynamic network topology. This covers dynamic network encoders.
\item Section \ref{sec:dynamiclinkprediction} is an overview of how the encoders from section \ref{sec:deeplearningmodels} are used for prediction. This includes dynamic network decoders, loss functions and evaluation metrics.
\end{enumerate}

\section{Dynamic networks}
\label{sec:org0ff799d}
\label{sec:dynnetrepresentation}
A complex network is a representation of a complex system. A network that changes over time can be represented as a dynamic network. A dynamic network has both temporal and structural patterns, and these patterns are described by a dynamic network model. 

The definition of a link is essential to any network representation. It is even more essential in dynamic networks, as it dictates when a link appears and disappears. Different link definitions affect network properties which in turn affect which models are capable of representing the dynamic network.

Dynamic networks are complex networks that change over time. Links and nodes may appear and disappear. With only this insight we can form a general definition for dynamic networks. Our definition is inspired by Rossetti and Cazabet \cite{rossettiCommunityDiscoveryDynamic2018}.

\textbf{Definition 1 (Dynamic Network)} A Dynamic Network is a graph \(G = (V, E)\) where: \(V = \{(v, t_s, t_e)\}\), with \(v\) a vertex of the graph and \(t_{s}, t_{e}\) are respectively the start and end timestamps for the existence of the vertex (with t\textsubscript{s} \(\le\) t\textsubscript{e}). \(E = \{(u,v,t_s,t_e)\}\), with \(u,v \in V\) and \(t_{s}, t_{e}\) are respectively the start and end timestamps for the existence of the edge (with t\textsubscript{s} \(\le\) t\textsubscript{e}).

This definition and any of the later definitions represent unlabeled and undirected networks, but they can however trivially be extended with both direction and labels taken into account. 

Whereas dynamic networks are defined as complex networks where links and nodes may appear and disappear, dynamic network models are often designed to work on specific kinds of dynamic networks and specific dynamic network representations. It, therefore, makes sense to distinguish between different kinds of dynamic networks and how they are represented.

Table \ref{tab:notation} an overview of the notation and Table \ref{tab:abbreviation} is an overview of the abbreviations used in this work.

There are several surveys on dynamic network methods \cite{aggarwalEvolutionaryNetworkAnalysis2014,kimReviewDynamicNetwork2017,zhangSurveyStreamingAlgorithms2010,holmeTemporalNetworks2012,holmeModernTemporalNetwork2015,fritzTempusVolatHora2019,mcgregorGraphStreamAlgorithms2014,aouayModelingDynamicsSocial2014,rossettiCommunityDiscoveryDynamic2018,kazemiRelationalRepresentationLearning2019}. These surveys focus either on specific kinds of dynamic networks or on a specific discipline and limit the scope of the survey to models in that discipline. To the best of our knowledge there is no comprehensive survey of dynamic networks, nor does any dynamic network model survey present a complete foundation or framework for dynamic networks. The aim of this section is to set the stage for the dynamic graph neural network survey by creating a conceptual framework for dynamic networks with more precise terminology and to add context by giving an overview of methods used for modelling dynamic network topology.

\subsection{Dynamic network representations}
\label{sec:org47574f1}
\label{sec:temporalrepresentations}
Dynamic networks can be represented in different ways and there are advantages and disadvantages inherent to the different representation types. 

Dynamic network representations can be grouped into four distinct levels ordered by temporal granularity: (i) static, (ii) edge-weighted, (iii) discrete, and (iv) continuous networks \cite{rossettiSocialNetworkDynamics2015}.

\begin{figure}[htbp]
\centering
\includegraphics[width=\linewidth]{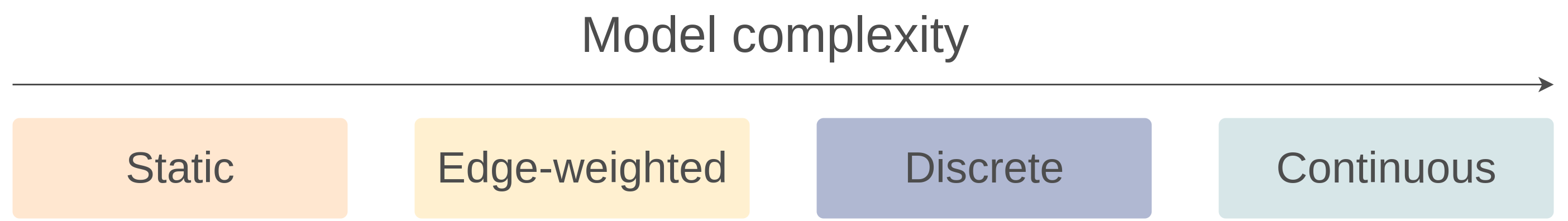}
\caption{\label{fig:dynamicnetworkrepresentation}
Network representations ordered by temporal granularity. Static networks are the most coarse-grained and continuous representations are the most fine-grained. With increasing temporal granularity comes increasing model complexity. The figure is inspired by Fig. 5.1 from Rossetti \cite{rossettiSocialNetworkDynamics2015}
}
\end{figure}

Fig. \ref{fig:dynamicnetworkrepresentation} shows those four representations with increasing model complexity as the model becomes more temporally fine-grained:
\begin{itemize}
\item \textbf{Static networks} have no temporal information.
\item \textbf{Edge weighted networks} have temporal information included as
  labels on the edges and/or nodes of a static network. The most straightforward example of this is a static network with the edges labelled with the time they were last active.
\item \textbf{Discrete networks} are represented in discrete time intervals. These can be represented by multiple snapshots of the network at different time intervals.
\item \textbf{Continuous networks} have no temporal aggregation applied to them. This   representation carries the most information but is also the most complex.
\end{itemize}

Static and edge-weighted networks are used to model stable patterns or the actual state of the network, whereas discrete and continuous methods are used for more dynamic modelling \cite{rossettiCommunityDiscoveryDynamic2018}. This work focuses on dynamic networks and will therefore only cover discrete and continuous representations.

Fine-grained representations can be trivially aggregated to produce coarser representations. For example, links in a continuous representation can be aggregated into snapshots (or time-windows) which is a discrete representation. Any discrete representation can combine the snapshots, yielding an edge-weighted representation and any edge-weighted representation can discard the weights thus yielding a static network.

\subsubsection{Discrete Representation}
\label{sec:org7ed8ee6}
Discrete representations use an ordered set of graphs (snapshots) to represent a dynamic graph. 
\begin{equation}
DG = \{G^1, G^2, \dotsc, G^T\},
\end{equation}

where \(T\) is the number of snapshots. Discrete representations, often simply referred to as "snapshots" is common for dynamic networks \cite{aggarwalEvolutionaryNetworkAnalysis2014,masudaGuideTemporalNetworks2016,holmeTemporalNetworks2012}. Using a discrete representation of the dynamic network allows for the use of static network analysis  methods on each of the snapshots. Repeated use of the static methods on each snapshot can then collectively give insight into the network's dynamics.

There are other approaches that effectively use snapshots as well. Overlapping snapshots such as sliding time-windows \cite{datarMaintainingStreamStatistics2002} are also used in dynamic network analysis to have less radical change from one network snapshot to the next \cite{moriniRevealingEvolutionsDynamical2017}. Discrete dynamic networks need not be represented as an ordered set of graphs, they may also be represented as a multi-layered network \cite{boccalettiStructureDynamicsMultilayer2014} or as a tensor \cite{dunlavyTemporalLinkPrediction2011}.

\subsubsection{Continuous Representation}
\label{sec:org82e1be3}
Continuous network representations are the only representations that have exact temporal information. This makes them the most complex but also the representation with the most potential. We cover three continuous representations: (i) the event-based; (ii) the contact sequence; and (iii) the graph streams. The first two representations are taken from the temporal network literature and they are suitable for networks where links do not persist for long \cite{holmeTemporalNetworks2012,masudaGuideTemporalNetworks2016,holmeModernTemporalNetwork2015}. The third representation, i.e. the graph stream, is used in dynamic networks where edges persist for longer \cite{aggarwalEvolutionaryNetworkAnalysis2014}. The focus in these representations is on when edges are active, with no mention of change on nodes. 
All three representations are described in more detail below:

\begin{enumerate}
\item \textbf{The event-based representation} 
\label{sec:org5483781}
includes the time interval at which the edge on a graph is active \cite{masudaGuideTemporalNetworks2016}. An event is synonymous with a link in this case.
It is a representation for dynamic networks focusing on link duration. The network is given by a time-ordered list of events which include the time at which the event appeared and the duration of the event.

\begin{equation}
EB=\{(u_i, v_i, t_i, \Delta_i); i = 1, 2, \ldots\},
\end{equation}

where \(u_i\) and \(v_i\) is a node pair on which the \(i\)-th event occurs, \(t_i\) is the timestamp for when the event starts and \(\Delta_i\) is the duration of the event. This is very similar to, and serves the same purpose as, the interval graph \cite{holmeTemporalNetworks2012}. The difference is that the interval graph has the time at which the event ends while the event-based representation has the duration of the event.

\item \textbf{The contact sequence representation}
\label{sec:org7ded86f}
is a simplification of the event-based representation. In a contact, sequence the link is instantaneous and thus no link duration is provided.

\begin{equation}
CS=\{(u_i, v_i, t_i); i = 1, 2, \ldots\},
\end{equation}

It is common to consider event times in real systems instantaneous if the duration of the event is short or not important \cite{holmeTemporalNetworks2012,masudaGuideTemporalNetworks2016}. Examples of systems where this representation is suitable, include message networks such as text message and email networks.

\item \textbf{The graph stream representation}
\label{sec:orga2e5328}
is used to represent static graphs that are too large to fit in memory but can also be used as a representation of a dynamic network \cite{zhangSurveyStreamingAlgorithms2010}. It is similar to the event-based representation, however, it treats link appearance and link disappearance as separate events. 

\begin{equation}
GS=\left\{ e_1, e_2, \ldots\right\},
\end{equation}

where \(e_i=\left(u_i, v_i, t_i, \delta_i\right)\), and \(u_i\) and \(v_i\) is the node pair on which the \(i\)-th event occurs, \(t_i\) is the time at which the event occurs, and \(\delta_i \in \{-1, 1\}\) where \(-1\) represents an edge removal and \(1\) represents that an edge is added.

The original representation (used for large graphs) does not include timestamped information of \emph{when} an edge is added/removed \cite{zhangSurveyStreamingAlgorithms2010}. Timestamps will have to be added for retrieving temporal information.

Since graph streams are mostly used to circumvent hardware limitations rather than a limitation of network representations, we will not survey them in detail here. For a more in-depth discussion of the graph streams, we refer the interested reader to \cite{zhangSurveyStreamingAlgorithms2010,aggarwalEvolutionaryNetworkAnalysis2014,mcgregorGraphStreamAlgorithms2014}.
\end{enumerate}

Which of the above representations is suitable for the network depends on the link duration with the intricacies of link duration covered in the next section.

\subsection{Link duration spectrum}
\label{sec:org7c7fd60}
\label{sec:dynamicnetworktypes}
\label{sec:linkdynamics}
Dynamic networks go by many names and sometimes these names indicate specific types of dynamic networks. There is substantial literature on 'temporal networks' \cite{holmeTemporalNetworks2012,holmeModernTemporalNetwork2015,masudaGuideTemporalNetworks2016} which focuses on highly dynamic networks where links may represent events such as human interactions or a single email. On the other hand, there is also literature that refers to slowly evolving networks, where links represent persistent relations \cite{aggarwalEvolutionaryNetworkAnalysis2014}. To the best of our knowledge, there are only two works that take note of this distinction, Rossetti and Cazabet \cite{rossettiCommunityDiscoveryDynamic2018}, and Holme \cite{holmeModernTemporalNetwork2015}. 

Rossetti and Cazabet \cite{rossettiCommunityDiscoveryDynamic2018} refer to temporal interaction and relational networks (our temporal and evolving networks respectively), but they do not categorize or make a formal distinction between the different networks. 

Holme \cite{holmeModernTemporalNetwork2015}  suggests that temporal networks can be distinguished by two requirements: (i) The dynamics on the network being at the same or at a similar time scale as the dynamics of the network; and (ii) The dynamic network is non-trivial at any given time (an instantaneous snapshot yield little to no network structure). 

The distinction manifests itself in networks even when not considering dynamics on the networks, and this work is limited to the dynamics of the network. Therefore we distinguish temporal networks purely based on network topology. We use the second requirement noted by Holme \cite{holmeModernTemporalNetwork2015}. 

This work not only provides a way to distinguish between temporal networks and dynamic networks, but it also proposes a framework in which all networks of dynamic topology fit. We do this by introducing the link duration spectrum.

\begin{figure}[htbp]
\centering
\includegraphics[width=\linewidth]{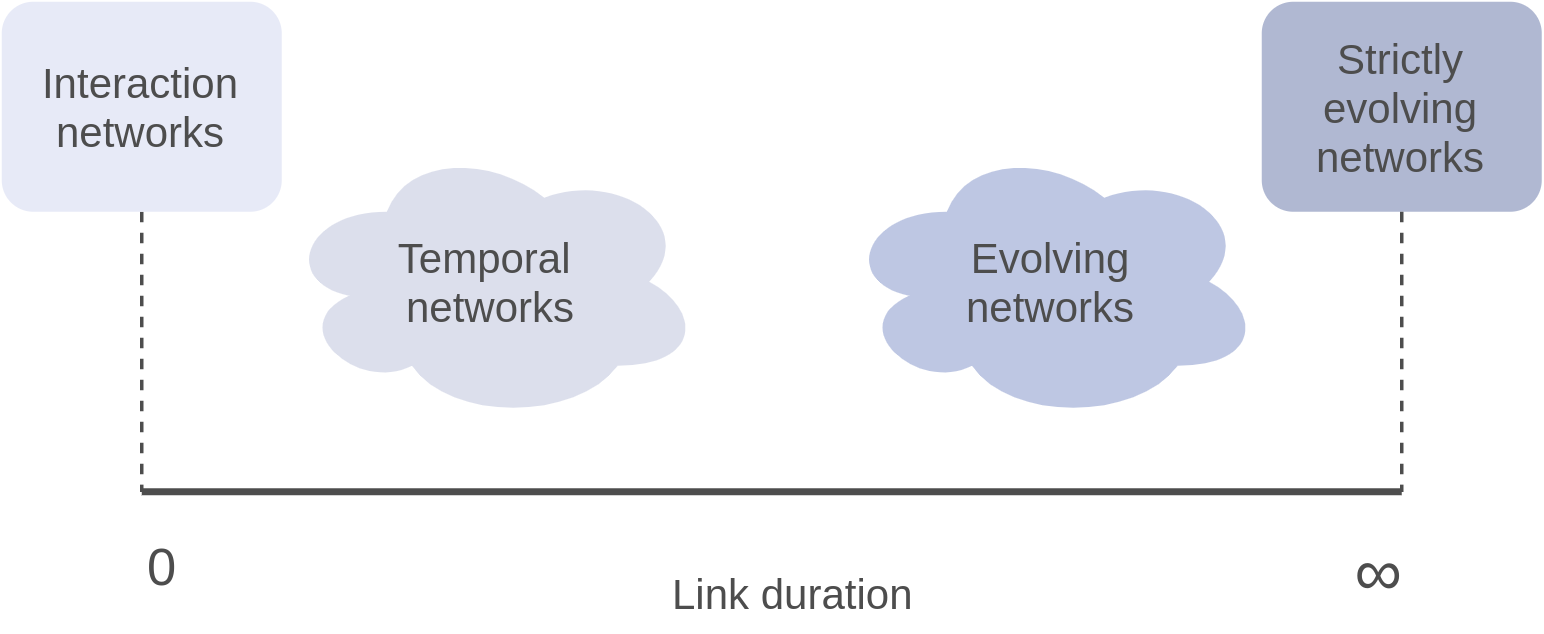}
\caption{\label{fig:eventdurationtypes}
Temporal and evolving networks on the link duration spectrum. The spectrum go from 0 (links have no duration) to infinity (links last forever).}
\end{figure}

Fig. \ref{fig:eventdurationtypes} shows different types of networks on the link duration spectrum. The scale goes from interactions with no link duration to links that have infinite link duration. No link ever disappears in a network with infinite link duration. Temporal networks reside on the lower end of the link duration spectrum, whereas evolving networks reside on the higher end. The distinction is as follows:
\begin{itemize}
\item \textbf{Temporal networks.} Highly dynamic networks which are too dynamic to be represented statically. The network is at any given time non-trivial. These networks are studied in the temporal network literature \cite{holmeTemporalNetworks2012,masudaGuideTemporalNetworks2016}. Network properties such as degree distribution and clustering coefficient cannot be adopted directly from static networks and are non-trivial to define. It is more natural to think of a link as an event with a duration.
\item \textbf{Evolving networks.} Dynamic networks where events persist for long enough to establish a network structure. An instantaneous snapshot yields a well-defined network. Network properties such as degree distribution and clustering coefficient can be adopted from static networks and gradually updated. These are the networks most often referred to when the term \emph{dynamic network} is used. Links persist for so long that it is more natural to think of link appearance as an event and link disappearance as another event.
\end{itemize}

Furthermore, there is one notable special case for each of the dynamic network types. These are types of networks that reside on the extreme ends of the link duration spectrum:
\begin{itemize}
\item \textbf{Interaction networks.} A type of temporal network where links are instantaneous events. These networks are studied in the temporal network literature and often represented as contact sequences \cite{holmeTemporalNetworks2012,masudaGuideTemporalNetworks2016}.
\item \textbf{Strictly evolving networks.} A type of evolving network where events have infinite duration. This implies that the links never disappear. 
\end{itemize}

\begin{figure}[htbp]
\centering
\includegraphics[width=\linewidth]{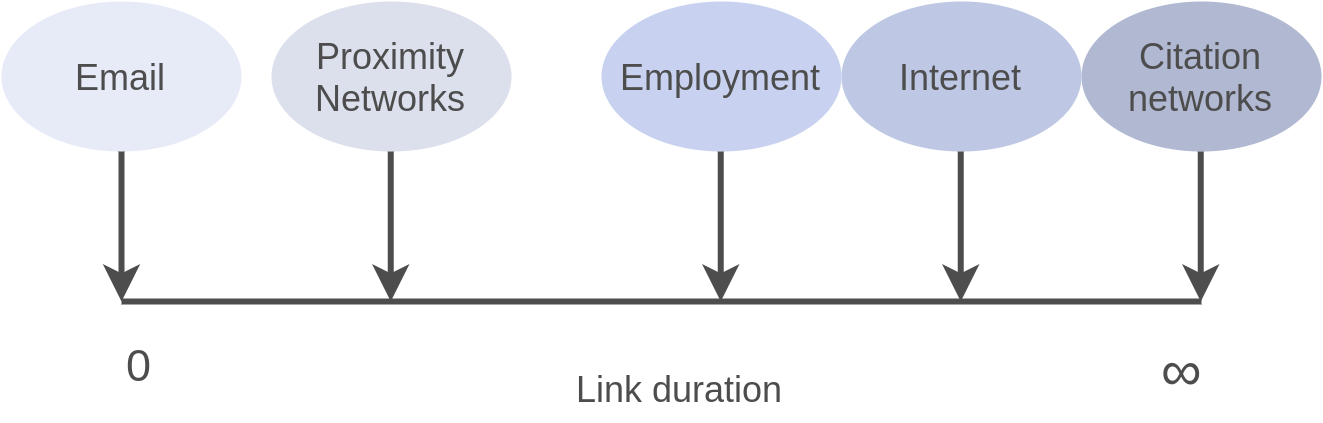}
\caption{\label{fig:eventdurationexample}
Examples of networks on the link duration spectrum.}
\end{figure}

Fig. \ref{fig:eventdurationexample} shows examples of networks on the link duration spectrum.
\begin{itemize}
\item An email is a nearly instantaneous event\footnote{If you model information propagation then in practice it takes time from the moment an email is sent until it is read, so that case considering the email an instantaneous event is an approximation.}, an email network can therefore be considered an \emph{interaction network}.
\item Proximity networks are used as an example of a temporal network in \cite{holmeTemporalNetworks2012}. The link is defined by who is close to whom at what time. Links require maintenance and do not typically last very long.
\item Employment networks are social networks where links are formed between employees and employers. The link requires an action after it has been established (termination of contract) to change its state, but also maintenance (continued work from the employee). This network resides in the fuzzy area between temporal and evolving networks and can be treated as either.
\item The Internet is an example of the network where we consider nodes linked if data-packets can flow between nodes. A link tends to persist for a long time once established and thus the internet can be thought of as an evolving network.
\item Citation networks where links are defined as one paper citing another have the most persisting links. Once a paper cites another paper, the link lasts forever. This leads to a strictly growing network where no edges disappear. These networks have the additional special characteristic that edges only appear when new nodes appear.
\end{itemize}

Link definitions influence link duration, which in turn influences a network type. Links can be modified in ways that alter their link duration (also known as time to live, TTL\cite{rossettiCommunityDiscoveryDynamic2018}). An email network could define a link as: Actors have once sent an email between each other. This would modify the email link, which is usually nearly instant in duration to a link that will never disappear. This modification moves the network all the way to the right on the spectrum shown in Fig. \ref{fig:eventdurationtypes}. It transforms an interaction network into a strictly evolving one. Another example of a modification is to use a time-window to force forgetting. A time-window can be applied to a citation network such that only citations which occurred during the time-window appear as links. This will move the network to the left on the link duration spectrum. Depending on the size of the time-window the modified network may be either an evolving or a temporal network.

An additional theoretical special case that is not covered by this concept is a network where links may only disappear. This special case may justify another dimension along which dynamic networks should be distinguished. 

\subsection{Node dynamics}
\label{sec:org75956fc}
\label{sec:nodedynamics}
Another distinguishing factor among dynamic networks is whether nodes may appear or disappear. When modelling networks, it is sometimes simpler to assume that the number of nodes may not change so that the only possible new links are links between already existing nodes. 

Many evolving network models assume that edges appear as a new node appears. These models include pseudo-dynamic models such as preferential attachment \cite{barabasiEmergenceScalingRandom1999}, forest fire \cite{leskovecGraphsTimeDensification2005} and GraphRNN \cite{youGraphRNNDeepGenerative2018}. This is fitting for a citation network where every node is a paper and the edges are cited papers, though, in many real-world networks, edges can appear and disappear regardless of whether nodes appear.

With respect to node change, we can distinguish between two kinds of networks.
\begin{itemize}
\item \textbf{Static} where the number of nodes stays static over time; and
\item \textbf{Dynamic} where the nodes may appear and disappear.
\end{itemize}

A notable special case of node-dynamic networks are the networks where nodes may only appear:
\begin{itemize}
\item \textbf{Growing} networks are those where nodes may only appear. We consider this a special case of node-dynamic networks.
\end{itemize}

We are unaware of any real-world networks where nodes may only disappear. But it should be noted as at least a theoretical special case. Node growing networks on the other hand are rather common. 

Any kind of node dynamics can be combined with any kind of link duration network. We can thus have, a \emph{growing evolving network} or a \emph{node-static temporal network}. Similarly to the edge duration spectrum, a node duration spectrum could theoretically be established, but it has no direct impact on dynamic network structure and we, therefore, chose to keep node dynamics a discrete distinction.

\begin{table}
\caption{\label{tab:dynamicnetworktypes}
Dynamic network types by node dynamics and link duration, excluding special cases.}
\adjustbox{max width=\linewidth}{
   \setlength{\extrarowheight}{2pt}
   \begin{tabular}{cc|c|c|}
     & \multicolumn{1}{c}{} & \multicolumn{2}{c}{Link duration}\\
     & \multicolumn{1}{c}{} & \multicolumn{1}{c}{Temporal}  & \multicolumn{1}{c}{Evolving} \\\cline{3-4}
     \multirow{2}*{Node dynamics}  & Static & Node-static temporal & Node-static evolving \\\cline{3-4}
     & Dynamic & Node-dynamic temporal & Node-dynamic evolving \\\cline{3-4}
   \end{tabular}
}
\end{table}

The node dynamics is an important consideration when modelling the network. Some models support node dynamics whereas others do not. 
\subsection{The dynamic network cube}

Many models assume that nodes disappear when there are no longer any links connected to such nodes. This scheme can work for evolving networks, but in temporal networks, it is common that nodes have no links for the majority of the time. Thus for a temporal network, it makes sense to model node dynamics separately from link dynamics.

\Figure[t][width=131mm]{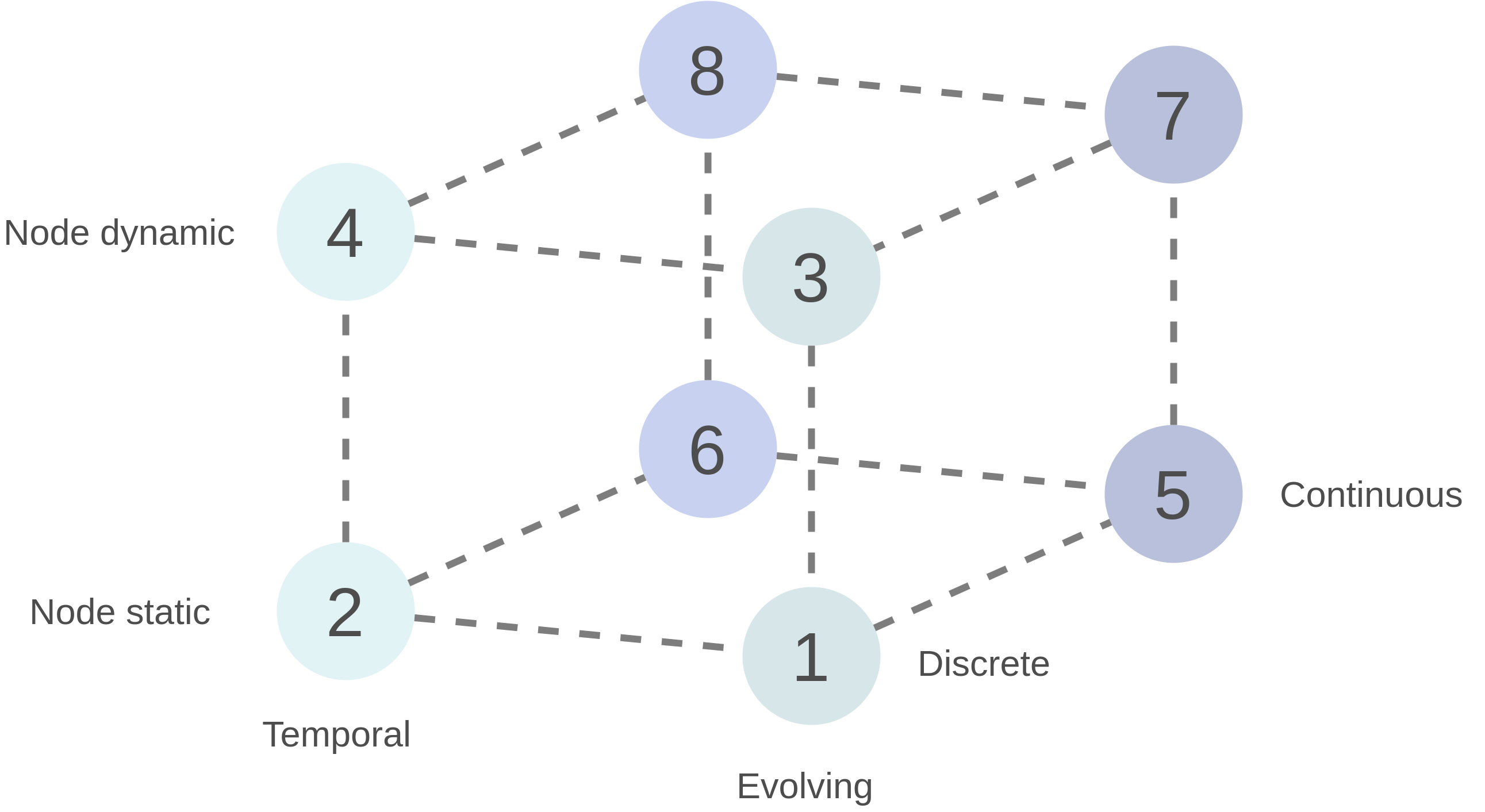}
{\label{fig:dynamicnetworkcube}
The dynamic network cube. The cube is a novel framework that succinctly represents different kinds of dynamic networks. Each node represents a specific type of dynamic networks. The nodes are organised along three dimensions: temporal granularity (discrete and continuous) from Section \ref{sec:temporalrepresentations}, the link duration spectrum (temporal and evolving) from Section \ref{sec:linkdynamics} and node dynamics (node-dynamic and node-static) from Section \ref{sec:nodedynamics}. The complete list of terminology from the cube is presented in Table \ref{tab:dynamicnetworkcube}.} 

\begin{table*}
\caption{Terminology of the dynamic network cube.}
\label{tab:dynamicnetworkcube}
\centering
\adjustbox{max width=\linewidth}{
\begin{tabular}{rllll}
Node & Temporal granularity & Node dynamics & Link duration & Precise dynamic network term\\
\hline
1 & Discrete & Node-static & Evolving & Discrete node-static evolving network\\
2 &  &  & Temporal & Discrete node-static temporal network\\
3 &  & Node-dynamic & Evolving & Discrete node-dynamic evolving network\\
4 &  &  & Temporal & Discrete node-dynamic temporal network\\
5 & Continuous & Node-static & Evolving & Continuous node-static evolving network\\
6 &  &  & Temporal & Continuous node-static temporal network\\
7 &  & Node-dynamic & Evolving & Continuous node-dynamic evolving network\\
8 &  &  & Temporal & Continuous node-dynamic temporal network\\
\end{tabular}
}
\end{table*}

\label{sec:dynnetframework}
Different aspects of dynamic network representation have been covered in the previous sections. Section \ref{sec:temporalrepresentations} defined different dynamic representations ordered by temporal granularity, section \ref{sec:linkdynamics} defined network types by link duration and section \ref{sec:nodedynamics} defined network types by node dynamics. This section will consider these previous sections jointly and discuss how the different network types fit together.

\begin{table}
\begin{centering}
\caption{Types of dynamic networks along three dimensions. Static networks and edge-weighted networks are not dynamic networks, but they are included for completeness. If we exclude special cases, we are left with two elements in each dimension.}
\label{tab:networkrepresentationtypes}
\adjustbox{max width=\linewidth}{
\begin{tabular}{lc}
Dimension & Network types\\
\hline
Temporal granularity & Static, edge-weighted, discrete, continuous\\
Link duration & Interaction, temporal, evolving, strictly evolving\\
Node dynamics & \makecell{Node-static, node-dynamic,\\ node-appearing, node-disappearing}\\
\end{tabular}
}
\end{centering}
\end{table}

Table \ref{tab:networkrepresentationtypes} includes a comprehensive list of the different dynamic network types. The types are grouped by node dynamic, temporal granularity and link duration type. Types of networks in each group can generally be combined, thus we can have a continuous node-static temporal network. The three groups can be thought of as dimensions of a space where different points in the space would represent different types of dynamic networks.

The 3D network type space resulting from excluding special cases is visualised in Fig. \ref{fig:dynamicnetworkcube}. When excluding special cases there are two types of networks along each dimension. The nodes are organised along three dimensions: temporal granularity (discrete and continuous) from Section \ref{sec:temporalrepresentations}, the link duration spectrum (temporal and evolving) from Section \ref{sec:linkdynamics} and node dynamics (node-dynamic and node-static) from Section \ref{sec:nodedynamics}.

Additionally, Table \ref{tab:dynamicnetworkcube} presents the suggested terminology for each of the dynamic network types. The precise dynamic network term column show the suggested terms for the different network types. These eight types represent domain-independent types of dynamic networks.

\subsection{Dynamic network models}
\label{sec:org0858441}
\label{sec:networkmodels}
This brief discussion on dynamic network models is intended to give a high-level overview of the dynamic model landscape without discussing different kinds of models in detail. For a detailed discussion, we refer to dedicated works. The aim of this section, is to give the reader the background and context needed to navigate through the field of dynamic network models.

A network model may model a variety of different network characteristics or dynamics. In this work, we focus on models of dynamic network structure. Many models define rules for how links are established \cite{barabasiEmergenceScalingRandom1999,leskovecGraphsTimeDensification2005}. The rules are defined such that a network evolved with those rules express some desired features. These features are often observed in real-world networks and then included in models as a rule. The search for a good dynamic network model is thus also a search for accurate rules on link formation.

Network models might aim to replicate characteristics like node degree distribution or average shortest path between nodes \cite{newmanNetworks2018}. The models define probabilistic rules for how links form such that the emerging network has certain distributions of given characteristics observed in real-world networks \cite{newmanNetworks2018}. Some dynamic network models, particularly temporal network models, focus on temporal aspects. An example of a temporal characteristic is the distribution of inter-event times \cite{masudaGuideTemporalNetworks2016}.

\Figure[ht][width=171mm]{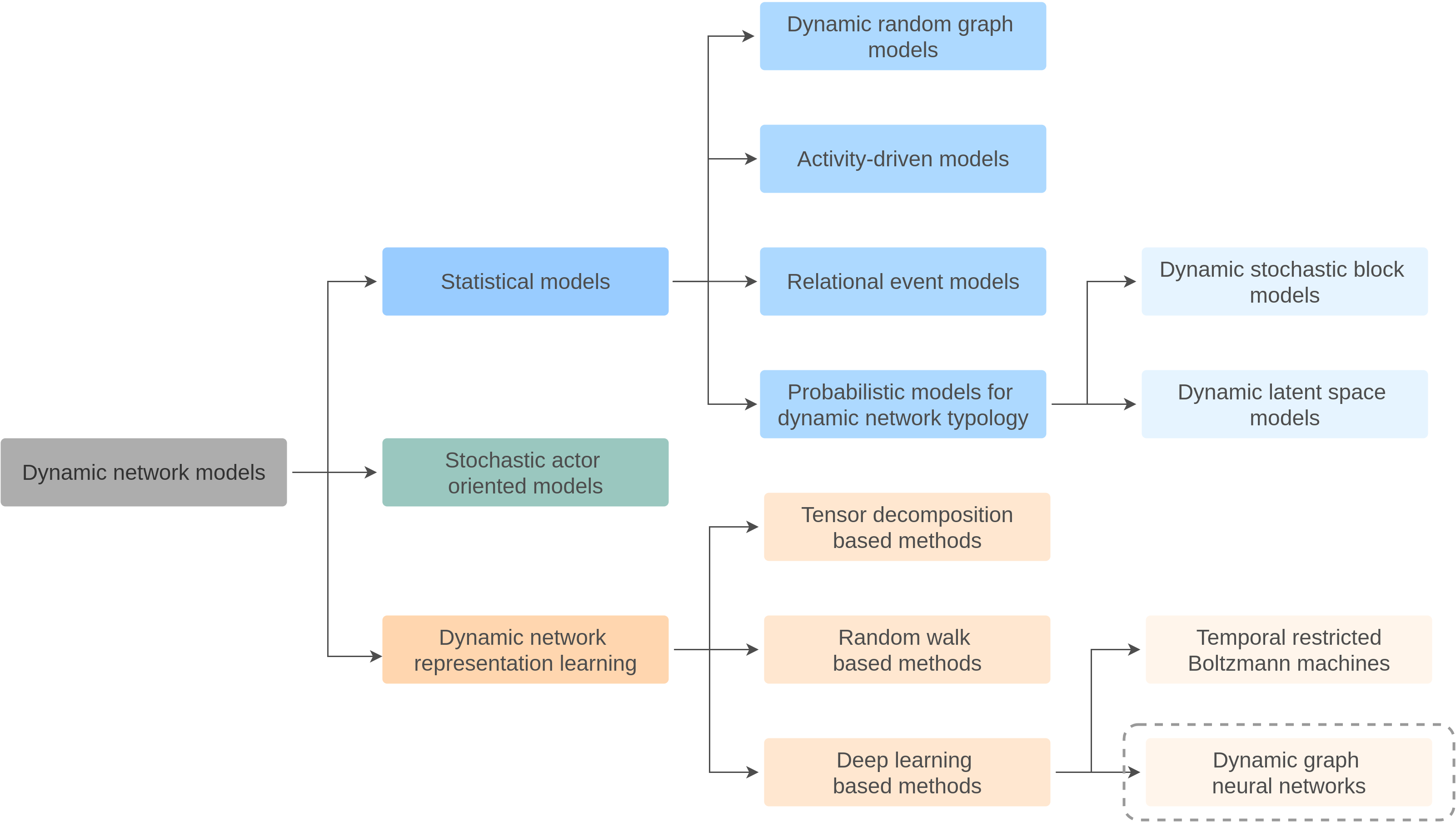} 
{\label{fig:dynamicnetworkmodels}
An overview of dynamic network models with dynamic graph neural networks outlined. Statistical models are models intended for inference or identifying statistical regularities in dynamic networks. Representation learning models are models which automatically detect features needed for the intended task. Stochastic actor oriented models are agent-based models. Dynamic network representation learning consist of shallow (tensor decomposition and random walk based) methods and deep learning based methods. This work explores dynamic graph neural networks in detail.}

There are several use cases for network models. They may be used as reference models \cite{holmeTemporalNetworks2012,holmeModernTemporalNetwork2015} or as realistic models \cite{barabasiEvolutionSocialNetwork2002, xu2015stochastic, chenELSTMDDeepLearning2019}, and depending on their purpose there are several tasks the model can be used for. These include:
\begin{itemize}
\item Reference models are used in the analysis of static networks to study the importance and role of structural features of static networks. Reference models aim to preserve some characteristic such as node degree distribution and otherwise create maximally random networks. The goal is to determine how the observed network is different from a completely random network with the same characteristics. This approach has been adapted to temporal networks \cite{holmeTemporalNetworks2012}. 
\item Realistic models aim to replicate the change in the network as closely as possible. They can be used for several tasks such as network prediction \cite{kazemiRelationalRepresentationLearning2019,chenELSTMDDeepLearning2019,trivedi2018dyrep} and community detection \cite{rossettiCommunityDiscoveryDynamic2018}. Examples include probabilistic models such as the dynamic stochastic block model \cite{xuDynamicStochasticBlockmodels2014} and representation learning based models such as E-LSTM-D \cite{chenELSTMDDeepLearning2019}. Some realistic models aim to generate (simulate) realistic networks \cite{youGraphRNNDeepGenerative2018,ashrafSimulationAugmentationSocial2019}. 
\end{itemize}

We establish a typology of models for dynamic network topology. The typology is based on the type of method used to model the network (see Fig. \ref{fig:dynamicnetworkmodels}).

We group models intended for inference or identifying statistical regularities under statistical models. These include dynamic random graph models, probabilistic models, activity driven models and relational event models. Random graph models (RGM) and Exponential random graph models (ERGM) are random graph models which produce randomly connected graphs while following known common network topology \cite{newmanNetworks2018}. Activity driven models are fit to interaction networks by modelling the activity of each node \cite{perra2012activity}. Relational event models are continuous-time models for interaction networks, they define the propensity for a future event to happen between node pairs.

Latent space models and stochastic block models are generative probabilistic models. Latent space models require the fitting of parameters with Markov chain Monte Carlo (MCMC) methods and are very flexible but scale to only a few hundred nodes \cite{junuthulaEvaluatingLinkPrediction2016}. Stochastic block models, on the other hand, scale to an order of magnitude larger networks, at a few thousand nodes \cite{junuthulaEvaluatingLinkPrediction2016}.

Stochastic actor oriented models (SAOM) are continuous-time models which consider each node an actor and model actor \emph{behaviour}. SAOMs learn to represent the dependencies between a network structure, the position of the actor and the actor behaviour \cite{snijdersIntroductionStochasticActorbased2010}. 

Dynamic network representation learning includes a diverse set of methods that can be used to embed the dynamic graph in a latent space. Representation learning on dynamic networks includes models based on tensor decomposition, random walks and deep learning. Since latent space models and stochastic block models also generate variables in a latent space they are closely related to dynamic network representation learning.

 Tensor decomposition is analogous to matrix factorization where the extra dimension is time \cite{kazemiRelationalRepresentationLearning2019}. Random walk approaches for dynamic graphs are generally extensions of random walk based embedding methods for static graphs or they apply temporal random walks \cite{masudaGuideTemporalNetworks2016}. Deep learning models include deep learning techniques to generate embeddings of the dynamic network. Deep models can be contrasted with the other networks representation learning models which are shallow models. We distinguish between two types of deep learning models: (i) Temporal restricted Boltzmann machines and (ii) Dynamic graph neural networks. Temporal restricted Boltzmann machines are probabilistic generative models which have been applied to the dynamic link prediction problem \cite{divakaran2019temporal,liDeepLearningApproach2014,liRestrictedBoltzmannMachineBased2018,liu2013deep}. Dynamic graph neural networks combine deep time series encoding with the aggregation of neighbouring nodes. Often discrete versions of these models take the form of a combination of a GNN and an RNN. Continuous versions of dynamic graph neural networks cannot make direct use of a GNN since a GNN require a static graph. Continuous DGNNs must therefore modify how node aggregation is done.

A detailed survey of all kinds of dynamic network models is too broad a topic to cover in detail by one survey. Deep learning based models for dynamic networks is a rapidly growing and exciting field, however, no existing survey focuses exclusively on dynamic graph neural networks (Kazemi \emph{et al.} \cite{kazemiRelationalRepresentationLearning2019}, Xie \emph{et al.} \cite{xie2020survey} and Barros \emph{et al.} \cite{barros2021survey} being the closest). 

 For the models not discussed in section \ref{sec:deeplearningmodels} there are several works describing and discussing them in detail. Random reference models for temporal networks are surveyed in \cite{holmeTemporalNetworks2012} and \cite{holmeModernTemporalNetwork2015}. For activity-driven models see Perra \emph{et al.} \cite{perra2012activity} and for an introduction to the Relational Event Model (REM) see Butts \cite{Butts2017}. See Hanneke \emph{et al.} \cite{hannekeDiscreteTemporalModels2010} for Temporal ERGMs (TERGM) on discrete dynamic networks. Block \emph{et al.} \cite{blockChangeWeCan2018} provides a comparison of TERGM and SAOM. Fritz \emph{et al.} \cite{fritzTempusVolatHora2019} provide a comparison of a discrete-time model, based on the TERGM, and the Relational Event Model (REM), a continuous-time model. Goldenberg \emph{et al.} \cite{goldenbergSurveyStatisticalNetwork2010} survey dynamic network models and their survey include dynamic random graph models and probabilistic models. Kim \emph{et al.} \cite{kimReviewDynamicNetwork2017} surveys latent space models and stochastic block models for dynamic networks. For an introduction to SOAM see Snijders \emph{et al.} \cite{snijdersIntroductionStochasticActorbased2010}. For surveys of representation learning on dynamic networks see Kazemi \emph{et al.} \cite{kazemiRelationalRepresentationLearning2019}, Xie \emph{et al.} \cite{xie2020survey} and Barros \emph{et al.} \cite{barros2021survey}, and for a survey of dynamic link prediction, including Temporal restricted Boltzmann machines, see Divakaran \emph{et al.} \cite{divakaran2019temporal}.

\subsection{Discussion and summary}
\label{sec:org8d3e375}
\label{sec:dynnetsummary}
We have given a comprehensive overview of dynamic networks. This establishes a foundation on which dynamic network models can be defined and thus sets the stage for the survey on dynamic graph neural networks. Establishing this foundation included the introduction of a new taxonomy for dynamic networks and an overview of dynamic network models.

Section \ref{sec:temporalrepresentations} presents representations of dynamic networks and distinguishes between discrete and continuous dynamic networks.
In section \ref{sec:dynamicnetworktypes} we introduce the link duration spectrum and distinguish between temporal and evolving networks, and in section \ref{sec:nodedynamics} node dynamics is discussed, we distinguish between node-static and node-dynamic networks. Section \ref{sec:dynnetframework} brings together the previous sections to arrive at a comprehensive dynamic network taxonomy.

Discrete representations have seen great success in use on evolving networks with slow dynamics. Graph streams are used on evolving networks that update too frequently to be represented well by snapshots \cite{aggarwalEvolutionaryNetworkAnalysis2014}. Both discrete and continuous representations are used to represent temporal networks \cite{holmeTemporalNetworks2012,masudaGuideTemporalNetworks2016}. 
Table \ref{tab:taxonomysummary} combines information from section \ref{sec:temporalrepresentations} and section \ref{sec:linkdynamics} and summarizes the existing representations in terms of temporal granularity and link duration. 
\begin{table}[htbp]
\caption{\label{tab:taxonomysummary}
Suitable dynamic network representations for temporal and evolving networks.}
\adjustbox{max width=\linewidth}{
\centering
\begin{tabular}{c|cc}
Temporal granularity & Temporal network & Evolving network\\
\hline
Continuous & \makecell{Event-based representation\\ or Contact sequence} & Graph stream\\[0.25cm]
Discrete & Time-windows & Snapshots\\
\end{tabular}
}
\end{table}

Discrete representations have several advantages. A model which works on the static network case can be extended to dynamic networks by applying it on each snapshot and then aggregating the results of the model \cite{kimReviewDynamicNetwork2017, kazemiRelationalRepresentationLearning2019}. This makes it relatively easy, compared to the continuous representation to design dynamic network models. Furthermore, the distinction between an evolving and a temporal network is less important. If modelling a temporal network, one only needs to make sure that a time-window size is large enough that the network structure emerges in each snapshot. However, the discrete representations have their disadvantages too. Chief among them is coarse-grained temporal granularity. When modelling a temporal network the use of a time-window is a must. By using a time-window the appearance order of the links and temporal clustering (links appearing frequently together) is lost. 

Reducing the size of the time-window or the interval between snapshots is a way to increase temporal granularity. There are however some fundamental problems with this. In the case of a temporal network, a small time-window will eventually yield a snapshot with no network structure. In the case of an evolving network, we will have a sensible network no matter how small the time-window, however, there is a trade-off with run-time complexity. Discrete models tend to process the entire graph in each snapshot. In which case the run-time will increase linearly with the number of snapshots. The run-time problem is exacerbated by the fact that a lot of real-world graphs are huge which make the run-time on each snapshot significant. 

Continuous representations offer superior temporal granularity and thus theoretically a higher potential to model dynamic networks. However, continuous-time models tend to be more complex and require either completely new models or significant changes to existing ones to work on the continuous representation. Continuous models are less common than discrete-time models \cite{aggarwalEvolutionaryNetworkAnalysis2014,rossettiCommunityDiscoveryDynamic2018,kazemiRelationalRepresentationLearning2019}. This is likely due to continuous methods being significantly more difficult to develop than discrete methods \cite{aggarwalEvolutionaryNetworkAnalysis2014}. 

When modelling dynamic networks in continuous time it is essential to specify \emph{which} kind of network is being modelled. As models for temporal and evolving networks may not be mutually exclusive and many models work on only specific types of networks. In these cases, it might be possible to modify the link duration of a network to run a model on the network. This modification may come at the loss of information, for example when modifying an interaction network to a strictly evolving network, any reappearing link will be removed.

This entire background section establishes a foundation and a conceptual framework in which dynamic networks can be understood. By providing an overview of dynamic network models, it maps out the landscape around deep learning on dynamic graphs thus providing the necessary context. The following sections will explore dynamic graph neural networks in detail.

\section{Dynamic graph neural networks}
\label{sec:deeplearningmodels}
Network representation learning and Graph Neural Networks (GNN) have seen rapid progress recently and they are becoming increasingly important in complex network analysis. Most of the progress has been done in the context of static networks, with some advances being extended to dynamic networks. Particularly GNNs have been used in a wide variety of disciplines such as chemistry \cite{kearnesMolecularGraphConvolutions2016, duvenaudConvolutionalNetworksGraphs2015}, recommender systems \cite{ying2018graph, monti2017geometric} and social networks \cite{qiuDeepInfSocialInfluence2018, liu2019characterizing}.

GNNs are deep neural network architectures that encode graph structures. They do this by aggregating features of neighbouring nodes together. One might think of this node aggregation as similar to the convolution of pixels in convolutional neural networks (CNN). By aggregating features of neighbouring nodes together GNNs can learn to encode both local and global structure.

\Figure[ht][width=131mm]{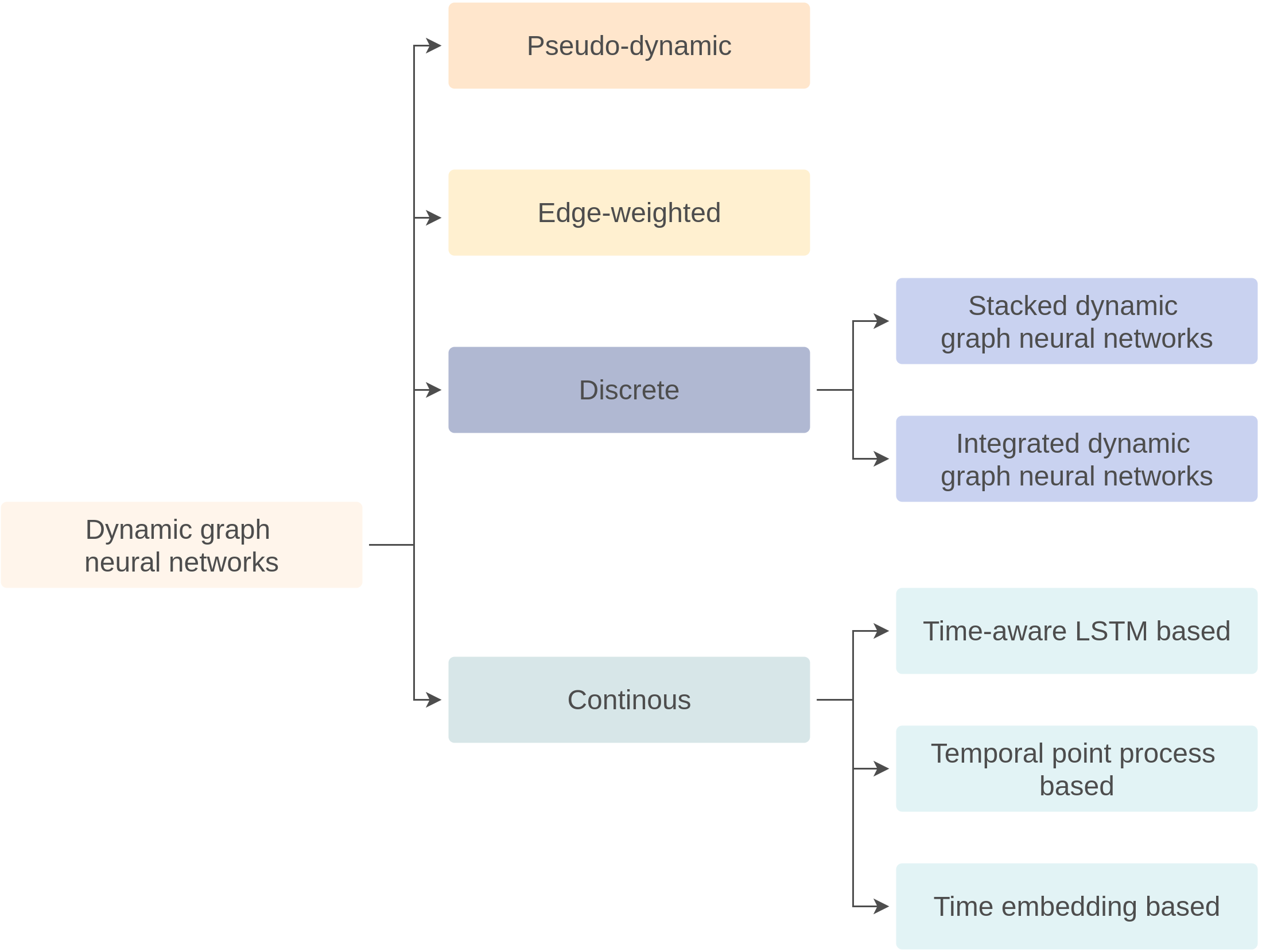}
{\label{fig:dgnnmodels}
An overview of the different types of dynamic graph neural networks. This is an extension of Fig \ref{fig:dynamicnetworkmodels} where we zoom in on graph neural networks. Different models are first grouped by which type of network they encode (pseudo-dynamic, edge-weighted, discrete or continuous). Discrete models are grouped by whether the structural layers and temporal layers are stacked, or integrated into one layer. Continuous models are grouped by how they encode temporal patterns.
}

Several surveys exist of works on static graph representation learning \cite{hamiltonInductiveRepresentationLearning2017,goyalGraphEmbeddingTechniques2018} and static graph neural networks \cite{zhouGraphNeuralNetworks2018,zhangDeepLearningGraphs2018,wuComprehensiveSurveyGraph2019}. Time-series analysis is relevant for work on dynamic graphs, thus recent advances in this domain is of relevance. For and up to date survey of deep learning on time series we refer to Fawaz \emph{et al.} \cite{fawazDeepLearningTime2019}. 

If dealing with an evolving graph, a static graph algorithm can be used to maintain a model of the graph. Minor changes to the graph would most likely not change the predictions of a static model too much, and the model can then be updated at regular intervals to avoid getting too outdated. We suspect that a spatial GNN is likely to stay accurate for longer than a spectral GNN, since the spectral graph convolution is based on the graph laplacian which will go through more changes than the local changes in a spatial GNN.

It is important to define what we mean by a dynamic graph neural network (DGNN). Informally we can say that a DGNN is a neural network that encodes a dynamic graph. However, there are some representation learning models for dynamic graphs using deep methods, which we do not consider dynamic graph neural networks. A key characteristic of a graph neural network is an aggregation of neighbouring node features (also known as message passing) \cite{zhouGraphNeuralNetworks2018}. Thus, if a deep representation learning model aggregates neighbouring nodes as part of its neural architecture we call it a dynamic graph neural network. In the discrete case, a DGNN is a combination of a GNN and a time series model. Whereas in the continuous case we have more variety since the node aggregation can no longer be done using traditional GNNs. 
Given this definition of representation learning, network models where RNNs are used but network structure is learned using other methods than node aggregation (temporal random walks for example), are not considered DGNNs.

The previous section (Section \ref{sec:dynnetrepresentation}) introduced a framework for dynamic networks and an overview of dynamic network models. The overview presented in Fig. \ref{fig:dynamicnetworkmodels} shows dynamic graph neural networks to be a part of deep representation learning, which in turn is part of dynamic network representation learning. We further extend the overview in Fig. \ref{fig:dynamicnetworkmodels} to show a hierarchical overview of dynamic graph neural networks, Fig. \ref{fig:dgnnmodels}. 

An overview of the types of DGNN encoders is seen in Fig. \ref{fig:dgnnmodels}. The encoders are grouped first by which type of network they encode, then by model type. The pseudo-dynamic approaches model a network with changing topology, but not time. Discrete DGNNs model discrete networks and continuous DGNNs model continuous networks. A discrete DGNNs encode the network snapshot by snapshot and encode a snapshot all at once, similar to how a GNN encode a static network. A continuous DGNN iterate over the network edge by edge and is thus completely independent of any snapshot size.

Common to all DGNNs is that the encoders aim to capture both structural and temporal patterns and store these patterns in embeddings. A stacked DGNNs separate encoding of structural and temporal patterns in separate layers, having one layer for structural patterns (using a static GNN) and one layer for temporal patterns (often using some form of an RNN), these models often make use of existing layers and combine them in new ways to encode dynamic networks. Integrated DGNNs combine structural and temporal patterns in one layer. This means that integrated DGNNs require the design of new layers, not just a combination of existing layers. The continuous DGNNs consist of RNN, Temporal point process (TPP) and time embedding based methods.

\Figure[ht][width=175mm]{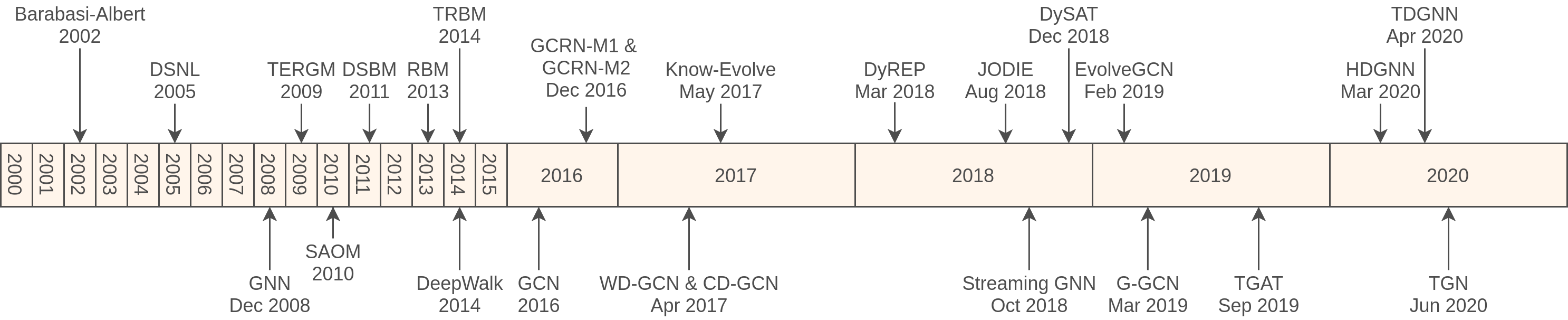}
{\label{fig:timeline}
Timeline of dynamic graph models and dynamic graph neural networks. The timeline shows the first dynamic network models of each type of model from Fig \ref{fig:dynamicnetworkmodels} and significant representation learning models leading up to the first DGNN. After the first DGNNs (GCRN-M1 and GCRN-M2\cite{seoStructuredSequenceModeling2018}) in Dec 2016, only DGNNs are marked on the timeline. DGNNs are marked by the month they were first publicised as they appeared in tight succession. The timeline indicates when a model was first publicized (the timeline may therefore show a different year than that in the citation if the paper was pre-published)
}

A timeline of dynamic network models with a focus on DGNNs is shown in  Fig. \ref{fig:timeline}. The timeline includes the first appearance of each of the models found in Fig. \ref{fig:dynamicnetworkmodels}, significant network embedding models preceding DGNNs and DGNNs. 

We consider the Albert-Barabasi model \cite{barabasiEvolutionSocialNetwork2002} the first dynamic network model, although it is only a pseudo-dynamic model (see section \ref{sec:deeppseudo}). The Dynamic Social Network in Latent space" (DSNL) model \cite{sarkar2005dynamic} is the first dynamic latent space model \cite{kimReviewDynamicNetwork2017}. The Temporal Exponential Random Graph Model (TERGM) \cite{hannekeDiscreteTemporalModels2010} a type of dynamic random graph model was introduced in 2009. Snijders \emph{et al.} introduced Stochastic Actor Oriented Models (SAOM) \cite{snijdersIntroductionStochasticActorbased2010} for dynamic networks in 2010. The first dynamic stochastic block model (DSBM) was introduced by Yang \emph{et al.} \cite{yang2011detecting}. The first restricted boltzmann machine (RBM) for static social networks \cite{liu2013deep} in 2013 was shortly followed by the first RBM for dynamic networks, the Temporal Restricted Boltzmann Machine (TRBM) in 2014.

Prior to DGNNs there were several influential static embedding methods and graph neural networks. The first GNN \cite{scarselli2008graph} was introduced in 2008. Deepwalk \cite{perozziDeepWalkOnlineLearning2014}, a highly influential node embedding fueled by random walks was introduced in 2014. Some Graph Convolutional Neural networks (GCN) \cite{defferrardConvolutionalNeuralNetworks2016,kipfSemisupervisedClassificationGraph2016} which function as building blocks and inspiration for several DGNNs were released in 2016.

The first DGNNs were discrete DGNNs. First (GCRN-M1 \& GCRN-M2) was introduced by Seo \emph{et al.} \cite{seoStructuredSequenceModeling2018}, followed by Manessi \emph{et al.} \cite{manessiDynamicGraphConvolutional2020} a few months later. Know-Evolve \cite{trivediKnowEvolveDeepTemporal2017} a TPP based model was the first continuous model, which in turn directly inspired DyREP \cite{trivedi2018dyrep} by the same author. JODIE \cite{kumar2019predicting} is notable as the RNN based DGNN, and it was quickly followed by Streaming GNN \cite{maStreamingGraphNeural2018} which was the first DGNN for continuous strictly evolving networks. DySAT \cite{sankarDynamicGraphRepresentation2018} introduced the first discrete DGNN which was based solely on attention, thus not using an RNN. EvolveGCN \cite{parejaEvolveGCNEvolvingGraph2019} introduced the first design that had an RNN feed into a GCN, rather than what the previous models did, which was to have a GCN feed into an RNN. The first pseudo-dynamic GNN, G-GCN was introduced in early 2019. TGAT \cite{xu2020inductive} is the first DGNN to encode inter-event time as a vector, while TGN \cite{rossi2020temporal} adds a memory module to TGAT. HDGNN showed how to use DGNNs for encoding discrete heterogeneous dynamic networks and TDGNN although simple was the first GNN to explicitly weight the edges to enable interaction network encoding.

This section surveys DGNNs, identifies different types of DGNNs and covers how embeddings are encoded. The next section (Section \ref{sec:dynamiclinkprediction}) covers decoding of the embeddings.

\subsection{Pseudo-dynamic models}
\label{sec:orgfa469fe}
\label{sec:deeppseudo}
Goldenberg \emph{et al.} \cite{goldenbergSurveyStatisticalNetwork2010} refer to network models as "pseudo-dynamic" when they contain dynamic processes, but the dynamic properties of the model are not fit to the dynamic data.  A well-known example of a non-DGNN pseudo-dynamic model is the Barabasi-Albert model \cite{barabasiEvolutionSocialNetwork2002}. 

G-GCN \cite{xuGenerativeGraphConvolutional2019} can be seen as an extension of the Variational Graph Autoencoder (VGAE) \cite{kipfVariationalGraphAutoencoders2016} which is able to predict links for nodes with no prior connections, the so-called cold start problem. It uses the same encoder and decoder as VGAE, namely a GCN \cite{kipfSemisupervisedClassificationGraph2016} for encoding and the inner product between node embeddings as a decoder. The resulting model learns to predict links of nodes that have only just appeared.

\subsection{Edge-weighted models}
As noted earlier in Section \ref{sec:temporalrepresentations}, dynamic network representations can be simplified. One way to simplify the modelling is to convert the dynamic network to an edge-weighted network and then use a static GNN on the edge-weighted network. This is exactly what Temporal Dependent GNN (TDGNN) does \cite{qu_continuous-time_2020}. They convert an interaction network to an edge weighted network by using an exponential distribution. An edge which appeared more recently gets a high weight and one that appeared long ago gets a low weight. After the conversion an standard GCN \cite{kipfSemisupervisedClassificationGraph2016} is applied to the edge-weighted network.  While the conversion from interaction network (a continuous network) to edge-weighted is done as part of the model in the original work, there appears to be is no reason why it cannot be done as a pre-processing step and thus we classify it as an edge-weighted model.

\subsection{Discrete Dynamic Graph Neural Networks}
\label{sec:orga2567f8}
Modelling using discrete graphs has the advantage that static graph models can be used on each snapshot of the graph. Discrete DGNNs use a GNN to encode each graph snapshot. We identify two kinds of discrete DGNNs: Stacked DGNNs and Integrated DGNNs.

Autoencoders use either static graph encoders or DGNN encoders, however since they are trained a little differently from DGNNs and generally make use of (and thus extend) a DGNN encoder they are here distinguished from other models. 

A discrete DGNN combines some form of deep time-series modelling with a GNN. The time-series model often comes in the form of an RNN, but self-attention has also been used.

Given a discrete graph \(DG = \{G^1, G^2, \dotsc , G^T\}\) a discrete DGNN using a function \(f\) for temporal modelling can be expressed as:
\begin{equation}
\begin{aligned}
z_1^t, \ldots, z_{n}^{t} &=\operatorname{GNN}\left(G^{t}\right) \\
h_j^t &=f\left(h_j^{t-1}, z_j^{t}\right) \text { for } j \in[1, n]
\end{aligned}
\end{equation}

where \(f\) is a neural architecture for temporal modelling (in the methods surveyed \(f\) is almost always an RNN but can also be self-attention), \(z_i^t \in \mathbb{R}^{l}\) is the vector representation of node \(i\) at time \(t\) produced by the GNN, where \(l\) is the output dimension of the GNN. Similarity \(h_i^t \in \mathbb{R}^{k}\) is the vector representation produced by \(f\), where \(k\) is the output dimension of \(f\).

This can also be written as:
\begin{equation}
\begin{aligned}
Z^{t} &=\operatorname{GNN}\left(G^{t}\right) \\ H^{t} &=f\left(H^{t-1}, Z^{t}\right)
\end{aligned}
\end{equation}

Informally we can say that the GNN is used to encode each network snapshot and \(f\) (the RNN or self-attention) encodes across the snapshots.

Seo \emph{et al.} \cite{seoStructuredSequenceModeling2018} introduce two deep learning models which encode a static graph with dynamically changing attributes. Whereas the modelling of this kind of graph is outside the scope of the survey, the two models they introduced are, to the best of our knowledge, the first DGNNs. They introduce both a stacked DGNN and an integrated DGNN: (i) Graph Convolutional Recurrent Network Model 1 (GCRN-M1) and (ii) GCRN model 2 (GCRN-M2) respectively. Very similar encoders have been used in later publications for dynamic graphs.

\subsubsection{Stacked Dynamic Graph Neural Networks}
\label{sec:org8df7a60}
The most straightforward way to model a discrete dynamic graph is to have a separate GNN handle each snapshot of the graph and feed the output of each GNN to a time series component, such as an RNN. We refer to a structure like this as a stacked DGNN.

There are several works using this architecture with different kinds of GNNs and different kinds of RNNs. We'll use GCRN-M1 \cite{seoStructuredSequenceModeling2018} as an example of a stacked DGNN. This model stacks the spectral GCN from \cite{defferrardConvolutionalNeuralNetworks2016} and a standard peephole LSTM \cite{gersLearningPreciseTiming}:

\begin{equation}
\begin{aligned}
z_{t} &=\operatorname{GNN}\left(X_{t}\right) \\
i &=\sigma\left(W_i z_{t}+U_{i} h_{t-1}+w_{i} \odot c_{t-1}+b_{i}\right) \\
f &=\sigma\left(W_f z_{t}+U_{f} h_{t-1}+w_{f} \odot c_{t-1}+b_{f}\right) \\
c_{t} &=f_{t} \odot c_{t-1}\\
&\hspace{3.5mm}+i_{t} \odot \tanh \left(W_{c} z_{t}+U_{c} h_{t-1}+b_{c}\right) \\
o &=\sigma\left(W_{o} z_{t}+U_{o} h_{t-1}+w_{o} \odot c_{t}+b_{o}\right) \\
h_{t} &=o \odot \tanh \left(c_{t}\right) \end{aligned}
\end{equation}

Let \(X_t \in \mathbb{R}^{n \times d}\), \(W \in \mathbb{R}^{k \times nl}\), \(U \in \mathbb{R}^{k \times k}\) and \(h, w, c, b, i, f, o \in \mathbb{R}^{k}\). The gates which are normally vectors in the LSTM are now matrices. Also, \(z_t \in \mathbb{R}^{nl \times 1}\) is a vector and not a matrix. Even though the GNN used by Seo \emph{et al.} \cite{seoStructuredSequenceModeling2018} can output features with the same structure as the input, they reshaped the matrix into a vector. This allows them to use a one-dimensional LSTM to encode the entire dynamic network.

Whereas \cite{seoStructuredSequenceModeling2018} use a spectral GCN and a peephole LSTM this is not a limitation of the architecture as any GNN and RNN can be used. Other examples of stacked DGNNs are: RgCNN \cite{narayanLearningGraphDynamics2018} which use the Spatial GCN, PATCHY-SAN \cite{niepertLearningConvolutionalNeural2016} stacked with a standard LSTM and DyGGNN \cite{taheriLearningRepresentEvolution2019} which uses a gated graph neural network (GGNN) \cite{liGatedGraphSequence2017} combined with a standard LSTM.

\Figure[t!][width=151mm]{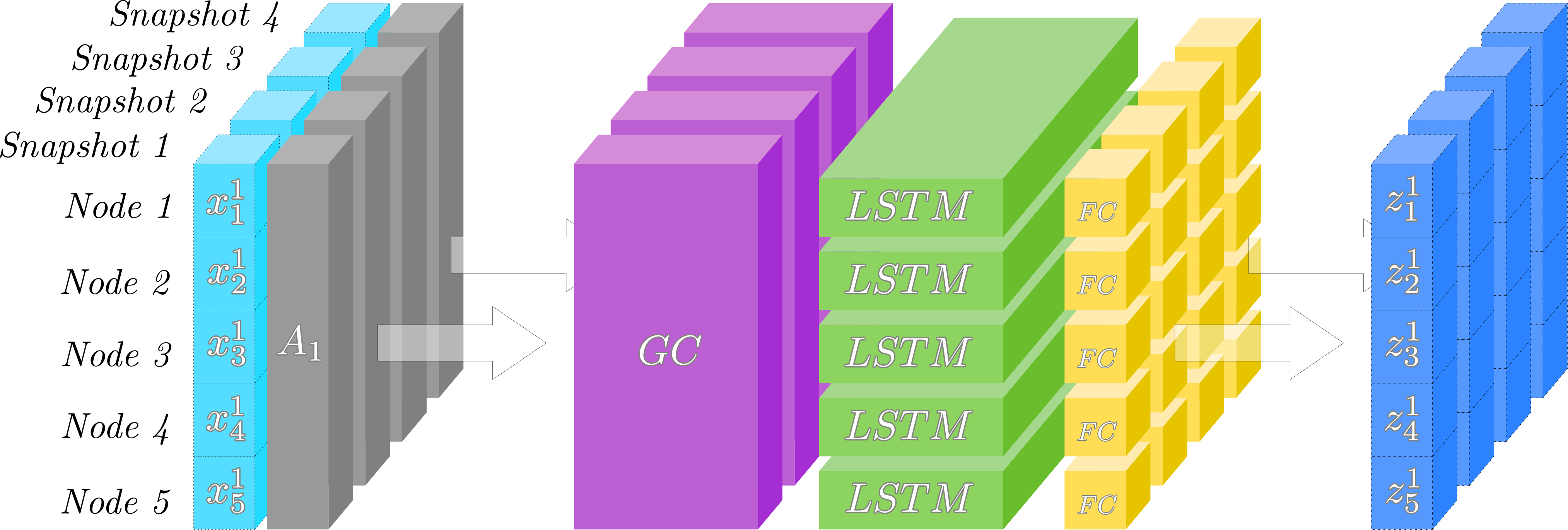} 
{\label{fig:stackedDGNN}
Stacked DGNN structure from Manessi \emph{et al.} \cite{manessiDynamicGraphConvolutional2020}. The graph convolution layer (GC) encode the graph structure in each snapshot while the LSTMs encode temporal patterns.}

Manessi \emph{et al.} \cite{manessiDynamicGraphConvolutional2020} present two stacked DGNN encoders: Waterfall Dynamic-GCN (WD-GCN) and Concatenated Dynamic-GCN (CD-GCN). These architectures are distinct in that they use a separate LSTM per node (although the weights across the LSTMs are shared). The GNN in this case is a GCN \cite{kipfSemisupervisedClassificationGraph2016} stacked with an LSTM per node. The WD-GCN encoder with a vertex level decoder is shown in Fig. \ref{fig:stackedDGNN}. WD-GCN and CD-GCN differ only in that CD-GCN adds skip-connections past the GCN. The equations below are for the WD-GCN encoder.

\begin{equation}
\begin{aligned}
Z_1, \dotsc, Z_t &= \operatorname{GNN}(A_1, X_1), \dotsc, \operatorname{GNN}(A_t, X_t) \\
H &= \operatorname{v-LSTM}_k(Z_1, \dotsc, Z_t)
\end{aligned}
\end{equation}

Let \(A \in \mathbb{R}^{n \times n}\) be the adjacency matrix, \(n\) be the number of nodes, \(d\) be the number of features per node and \(X_t \in \mathbb{R}^{n \times d}\) be the matrix describing the features of each node at time \(t\). \(Z_t \in \mathbb{R}^{n \times l}\) where \(l\) is the output size of the GNN and \(H \in \mathbb{R}^{k \times n \times t}\) where \(k\) is the output size of the LSTMs.
\begin{equation}
\begin{aligned}
\operatorname{v-LSTM}_k&(Z_1, \dotsc, Z_t)) =\\ 
&\left(
\begin{array}{c}
{\operatorname{LSTM}_{k}(V_{1}^{\prime} Z_1, \dotsc, V_{1}^{\prime} Z_t)} \\
{\vdots} \\
{\operatorname{LSTM}_{k}(V_{n}^{\prime} Z_1, \dotsc, V_{n}^{\prime} Z_t)} \\
\end{array}\right)
\end{aligned}
\end{equation}
where LSTM is a normal LSTM \cite{hochreiterLongShortTermMemory1997} and \(V_p \in \mathbb{R}^n\) is defined as \(V_p = \delta_{pi}\) where \(\delta\) is the Kronecker delta. Due to the v-LSTM layer the encoder can store a hidden representation per node.

Since a set of snapshots is a time-series, one is not restricted to the use of RNNs and other works have stacked GNNs with other types of deep time-series models. Sankar \emph{et al.} \cite{sankarDynamicGraphRepresentation2018} present a stacked architecture that consists completely of self-attention blocks. They use attention along the spatial and temporal dimensions. For the spatial dimension, they use the Graph Attention Network (GAT) \cite{velickovicGraphAttentionNetworks2017} and for the temporal dimension, they use a transformer \cite{vaswaniAttentionAllYou2017}. Wang \emph{et al.} \cite{wang2020generic,wang2020tedic} stacks a GNN with 1D temporal convolution (TNDCN) similar to the dilated convolution in WaveNet \cite{wavenet}.

Stacked DGNN architectures also exist for specific types of dynamic networks. There is HDGNN \cite{zhou2020heterogeneous} for heterogeneous dynamic networks and TeMP \cite{wu2020temp} for knowledge networks.

When encoding graphs one option is to split the graph into sub-graphs and use a GNN to project each sub-graph as done by Zhang \emph{et al.} \cite{zhang_link_2018} for static GNNs. This approach has also been applied to DGNNs by Cai \emph{et al.} \cite{cai2020structural}, where they split each snapshot into sub-graphs and use a stacked DGNN for anomaly detection. 

\subsubsection{Integrated Dynamic Graph Neural Networks}
\label{sec:orgc4bce7e}
Integrated DGNNs are encoders that combine GNNs and RNNs in one layer and thus combine modelling of the spatial and the temporal domain in that one layer.

Inspired by convLSTM \cite{shiConvolutionalLSTMNetwork2015} Seo \emph{et al.} \cite{seoStructuredSequenceModeling2018} introduced GCRN-M2. GCRN-M2 amounts to convLSTM where the convolutions are replaced by graph convolutions. ConvLSTM uses a 3D tensor as input whereas here we are using a two-dimensional signal since we have a feature vector for each node.

\begin{equation}
%
\begin{aligned}
f_t &=\sigma\left(W_f *_{\mathcal{G}} X_{t}+U_f *_{\mathcal{G}} h_{t-1}+ w_f \odot c_{t-1}+b_{f}\right) \\
i_t &=\sigma\left(W_i *_{\mathcal{G}} X_{t}+U_i *_{\mathcal{G}} h_{t-1}+ w_i \odot c_{t-1}+b_{i}\right) \\
c_t &=f_{t} \odot c_{t-1}\\
&\hspace{3.5mm}+i_{t} \odot \tanh \left(W_c *_{\mathcal{G}} X_{t}+U_c *_{\mathcal{G}} h_{t-1}+b_{c}\right) \\
o_t &=\sigma\left(W_o *_{\mathcal{G}} X_{t}+U_o *_{\mathcal{G}} H_{t-1}+ w_o \odot c_{t}+b_{o}\right) \\
h_{t} &=o \odot \tanh \left(c_{t}\right)
\end{aligned}
\end{equation}

where \(x_t \in \mathbb{R}^{n \times d}\), \(n\) is the number of nodes and \(x_i\) is a signal for the \emph{i}-th node at time \(t\).  \(W \in \mathbb{R}^{K \times k \times l}\) and \(U \in \mathbb{R}^{K \times k \times k}\) where \(k\) is the size of the hidden layer and \(K\) is the number of Chebyshev coefficients . \(W_f *_{\mathcal{G}} x_t\) denotes the graph convolution on \(x_t\).

\Figure[t!][width=151mm]{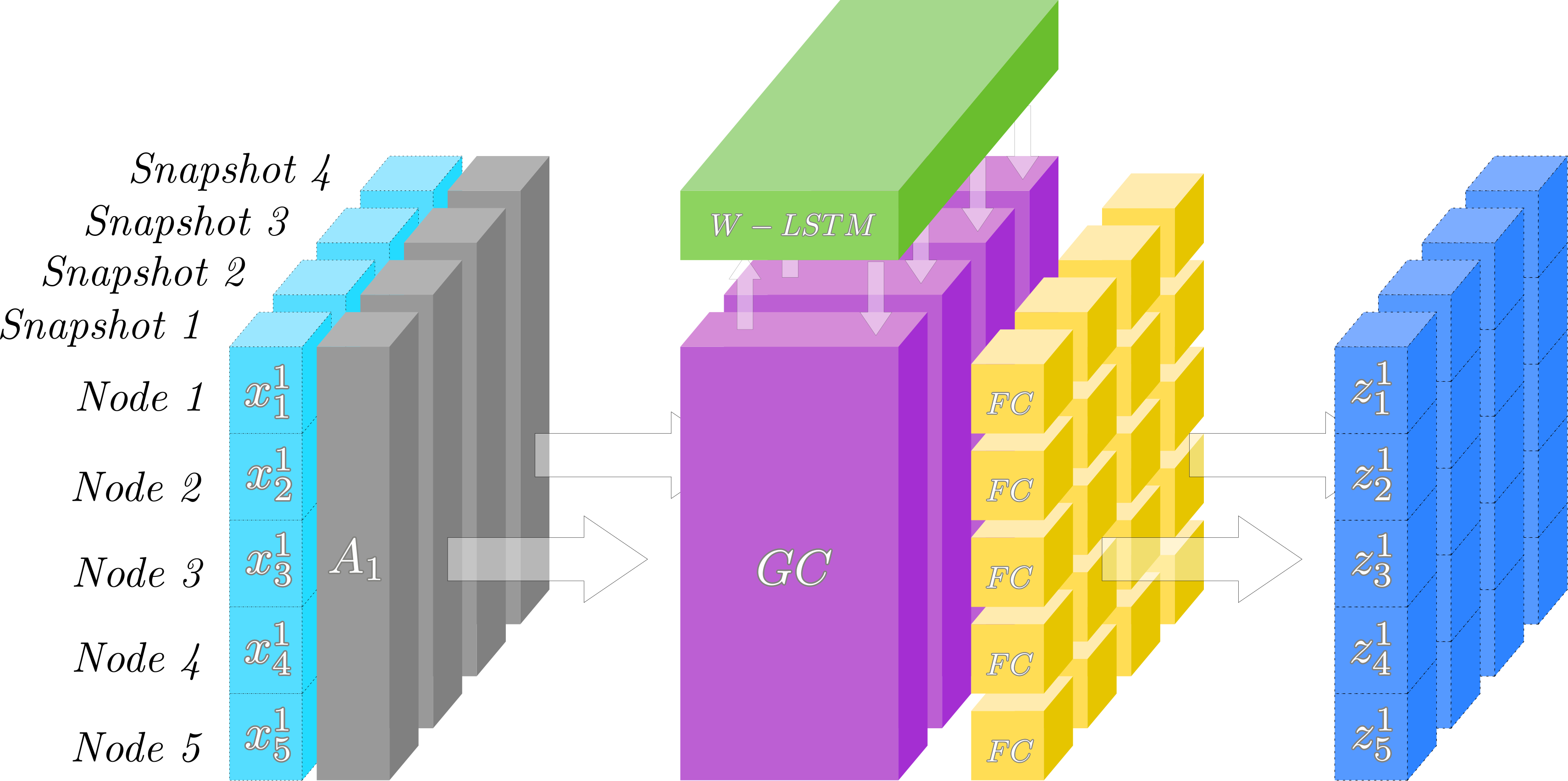} 
{\label{fig:integratedDGNN}
Integrated DGNN structure of EvolveGCN with an EGCU-O layer \cite{parejaEvolveGCNEvolvingGraph2019}. The EGCU-O layer constitutes the GC (graph convolution) and the W-LSTM (LSTM for GC weights). W-LSTM is used to initialize the weights of the GC.}

EvolveGCN \cite{parejaEvolveGCNEvolvingGraph2019} integrates an RNN into a GCN. The RNN is used to update the weights \(W\) of the GCN. \cite{parejaEvolveGCNEvolvingGraph2019} name their layer the Evolving Graph Convolution Unit (EGCU) and present two versions of it: (i) EGCU-H where the weights \(W\) are treated as the hidden layer of the RNN and (ii) EGCU-O where the weights \(W\) are treated as the input and output of the RNN. In both EGCU-H and EGCU-O, the RNN operate on matrices rather than vectors as in the standard LSTM. The EGCU-H layer is given by the following equations, where $(l)$ indicates the neural network layer:

\begin{equation}
\begin{aligned}
W_{t}^{(l)}&=\operatorname{GRU}\left(H_{t}^{(l)}, W_{t-1}^{(l)}\right) \\
H_{t}^{(l+1)}&=\operatorname{GNN}\left(A_{t}, H_{t}^{(l)}, W_{t}^{(l)}\right) \\
\end{aligned}
\end{equation}

And the EGCU-O layer is given by the equations:
\begin{equation}
\begin{aligned}
W_{t}^{(l)}&=\operatorname{LSTM}\left(W_{t-1}^{(l)}\right) \\
H_{t}^{(l+1)}&=\operatorname{GNN}\left(A_{t}, H_{t}^{(l)}, W_{t}^{(l)}\right) \\
\end{aligned}
\end{equation}
The RNN in both layers can be replaced with any other RNN, and the GCN \cite{kipfSemisupervisedClassificationGraph2016} can be replaced with any GNN given minor modifications.

Other integrated DGNN approaches are similar to GCRN-M2. They may differ in which GNN and/or which RNN they use, the target use case or even the kind of graph they are built for, but the structures of the neural architecture are similar. Examples of these include GC-LSTM \cite{chenGCLSTMGraphConvolution2018}, LRGCN \cite{liPredictingPathFailure2019}, RE-Net \cite{jin2019recurrent} and TNA \cite{bonner2019temporal}.

Chen \emph{et al.} \cite{chenGCLSTMGraphConvolution2018} present GC-LSTM, an encoder very similar to GCRN-M2. GC-LSTM takes the adjacency matrix \(A_t\) at a given time as an input to the LSTM and performs a spectral graph convolution \cite{defferrardConvolutionalNeuralNetworks2016} on the hidden layer. In contrast, GCRN-M2 runs a convolution on both the input and the hidden layer.

LRGCN \cite{liPredictingPathFailure2019} integrates an R-GCN \cite{schlichtkrullModelingRelationalData2018} into an LSTM as a step towards predicting path failure in dynamic graphs.

RE-Net \cite{jin2019recurrent} encodes a dynamic knowledge graph by integrating an R-GCN \cite{schlichtkrullModelingRelationalData2018} in several RNNs. Other modelling changes enable them to encode dynamic knowledge graphs, thus extending the use of discrete DGNNs to knowledge graphs.

A temporal neighbourhood aggregation (TNA) layer \cite{bonner2019temporal} stacks a GCN, a GRU and a linear layer. Bonner \emph{et al.} designs an encoder that stacks two TNA layers, to achieve a 2-hop convolution and employs variational sampling for use on link prediction. This architecture is arguably a stacked DGNN, but since the authors define the TNA as one layer, we classify it as an integrated DGNN, despite the layer itself being stacked.

\subsubsection{Dynamic graph autoencoders and generative models}
\label{sec:orge224437}
The Dynamic Graph Embedding model (DynGEM) \cite{goyal2018dynamicgem} uses a deep autoencoder to encode snapshots of discrete node-dynamic graphs. Inspired by an autoencoder for static graphs \cite{wangStructuralDeepNetwork2016} DynGEM makes some modifications to improve computation on dynamic graphs. The main idea is to have the autoencoder initialized with the weights from the previous snapshot. This speeds up computation significantly and makes the embeddings stable (i.e. no major changes from snapshot to snapshot). To handle new nodes the \emph{Net2WiderNet} and \emph{Net2DeeperNet} approaches from \cite{chen2015net2net} are used to add width and depth to the encoder and decoder while the embedding layer stays fixed in size. This allows the autoencoder to expand while approximately preserving the function the neural network is computing.

Dyngraph2vec \cite{goyalDyngraph2vecCapturingNetwork2019} is a continuation of the work done on DynGEM. dyngraph2vec considers the last \(l\) snapshots in the encoding and can thus be thought of as a sliding time-window. The adjacency matrices \(A^t, \dots , A^{t+l}\) are used to predict \(A^{t+l+1}\), it is assumed that no new nodes are added. The architecture comes in three variations: (1) dyngraph2vecAE, an autoencoder similar to DynGEM except that it leverages information from the past to make the future prediction; (2) dyngraph2vecRNN, where the encoder and decoder consist of stacked LSTMs; (3) dyngraph2vecAERNN, where the encoder has first a few dense feed-forward layers followed by LSTM layers and the decoder is similar to dyngraph2vecAE, namely a deep feed-forward network.

E-LSTM-D \cite{chenELSTMDDeepLearning2019} like DynGEM, encode and decode with dense layers, however, they run an LSTM on the encoded hidden vector to predict the new embeddings. Although trained like an autoencoder, the model aims to perform a dynamic link prediction.

Hajiramezanali \emph{et al.} \cite{hajiramezanaliVariationalGraphRecurrent2019} introduce two variational autoencoder versions for dynamic graphs: the Variational Graph Recurrent Neural Network (VGRNN) and Semi-implicit VGRNN (SI-VGRNN). They can operate on node-dynamic graphs. Both models use a GCN integrated into an RNN as an encoder (similar to GCRN-M2 \cite{seoStructuredSequenceModeling2018}) to keep track of the temporal evolution of the graph. VGRNN uses a VGAE \cite{kipfVariationalGraphAutoencoders2016} on each snapshot that is fed the hidden state of the RGNN \(h_{t-1}\). This is to help the VGAE take into account how the dynamic graph changed in the past. Each node is represented in the latent space and the decoding is done by taking the inner product decoder of the embeddings \cite{kipfVariationalGraphAutoencoders2016}. By integrating semi-implicit variational inference \cite{yinSemiImplicitVariationalInference2018} with VGRNN they create SI-VGRNN. Both models aim to improve dynamic link prediction.

Generative adversarial networks (GAN) \cite{goodfellowGenerativeAdversarialNets2014} have proven to be very successful in the computer vision field \cite{wang2019generative}. They have subsequently been adapted for dynamic network generation as well. GCN-GAN \cite{lei2019gcn} and DynGraphGAN \cite{xiong2019dyngraphgan} are two such models. Both models are aimed towards the dynamic link prediction task. The generator is used to generate an adjacency matrix and the discriminator tries to distinguish between the generated and the real adjacency matrix. The aim is to have the generator, generate realistic adjacency matrices which can be used as a prediction for the next time step.

GCN-GAN use a stacked DGNN as a generator and a dense feed-forward networks as a discriminator \cite{lei2019gcn} and DynGraphGAN use a shallow generator and a GCN \cite{kipfSemisupervisedClassificationGraph2016} stacked with a CNN as a discriminator \cite{xiong2019dyngraphgan}.

\subsection{Continuous Dynamic Graph Neural Networks}
\label{sec:orgc6489fe}
Currently, there are three DGNN approaches to continuous modelling. RNN based approaches where node embeddings are maintained by an RNN based architecture, temporal point based (TPP) approaches where temporal point processes are parameterized by a neural network and time embedding approaches where positional embedding of the time is used to represent time as a vector.

\subsubsection{RNN based models}
These models use RNNs to maintain node embeddings in a continuous fashion. A common characteristic for these models is that as soon as an event occurs or there is a change to the network, the embeddings of the interacting nodes are updated. This enables the embeddings to stay up to date continuously. There are two models in this category, Streaming graph neural networks (SGNN) \cite{maStreamingGraphNeural2018} which encode directed strictly evolving networks and JODIE \cite{kumar2019predicting} which encodes interaction networks.

The Streaming graph neural network \cite{maStreamingGraphNeural2018}  maintains a hidden representation in each node. The architecture consists of two components: (i) an update; and (ii) a propagation component. The update component is responsible for updating the state of the nodes involved in an interaction and the propagation component propagates the update to the involved node's neighbours.

The update and propagation component consist of 3 units each: (i) the interact unit; (ii) the update / propagate unit; and (iii) the merge unit. The difference between the update component and the propagation component is thus the second unit where the update component makes use of the update unit and the propagate component makes use of the propagate unit.

The model maintains several vectors for each node. Among them are: (i) a hidden state for the source role of the node; and (ii) a hidden state of the target role of the node. This is required to treat source and target nodes differently. The model also contains a hidden state which is based on both the source and target state of the node. The interact unit and merge units can be thought of as wrappers that handle many node states. The interact unit generates an encoding based on the interacting nodes and this can be thought of as an encoding of the interaction. The merge unit updates the combined hidden state of the nodes based on the change done to the source and target hidden states by the middle unit.

The middle units and core of the update and propagate components are the update and the propagate units. The update unit generates a new hidden state for the interacting nodes. It is based on a Time-aware LSTM \cite{baytasPatientSubtypingTimeAware2017}, which is a modified LSTM that works on time-series with irregular time intervals. The propagate unit updates the hidden states of the neighbouring nodes. It consists of an attention function \(f\), a time decay function \(g\) and a time based filter \(h\). \(f\) estimates the importance between nodes, \(g\) gauges the magnitude of the update based on how long ago it was and \(h\) is a binary function which filters out updates when the receiving node has too old information. \(h\) has the effect of removing noise as well as making the computation more efficient.

By first running the update component and afterwards propagating, information of the edge update is added to the hidden states of the local neighbourhood.

The second method is JODIE \cite{kumar2019predicting}. JODIE embeds nodes in an interaction network. It is however targeted towards recommender systems and built for user-item interaction networks. The intuition is that with minor modifications this model can work on general interaction networks.

JODIE uses an RNN architecture to maintain the embeddings of each node. With one RNN for users ($\operatorname{RNN}_u$) and one RNN for items ($\operatorname{RNN}_i$), the formula for each RNN is identical except that they use different weights. When an interaction happens between a user and an item, each of the embeddings is updated according to equation \ref{eqn:jodie}.

\begin{equation}
\label{eqn:jodie}
\resizebox{.8\hsize}{!}{
$
\begin{aligned}
u(t) &=\sigma\left(W_{1}^{u} u\left(\bar{t}\right)+W_{2}^{u} i\left(\bar{t}\right)+W_{3}^{u} f+W_{4}^{u} \Delta_{u}\right) \\
i(t) &=\sigma\left(W_{1}^{i} i\left(\bar{t}\right)+W_{2}^{i} u\left(\bar{t}\right)+W_{3}^{i} f+W_{4}^{i} \Delta_{i}\right)
\end{aligned}
$
}
\end{equation}

where $u(t)$ is the embedding of the interacting user, $i(t)$ the embedding of the interacting item, $u(\bar{t})$ the embedding of the user just before the interaction and similarly $i(\bar{t})$ is the embedding of the item just before the interaction. The superscript on the weights indicates which RNN they are parameters of, so $W_1^u$ is a parameter of $\operatorname{RNN}_u$. $f$ is the feature vector of the interaction  and $\Delta_u$ is the time since the user interacted with an item and similarly for $Delta_i$.

An additional functionality of JODIE is the projection component of their architecture. It is used to predict the trajectory of the dynamic embeddings. The model predicts the future position of the user or item embedding and is trained to improve this prediction. 

\subsubsection{Temporal point process based models}
Know-Evolve \cite{trivediKnowEvolveDeepTemporal2017} is the precursor to the rest of the dynamic graph temporal point process models discussed in this section. It models knowledge graphs in the form of interaction networks by parameterizing a temporal point process (TPP) by a modified RNN. With some minor modifications, the model should be applicable to any interaction network, but since the original model is specifically for knowledge graphs we will rather focus on its successor, DyREP \cite{trivedi2018dyrep}.

DyREP uses a temporal point process model which is parameterised by a recurrent architecture \cite{trivedi2018dyrep}. The temporal point process can express both dynamics "of the network" (structural evolution) and "on the network" (node communication). By modelling this co-evolution of both dynamics they achieve a richer representation than most embeddings.

The temporal point process (TPP) is modelled by events \((u, v, t, k)\) where \(u\) and \(v\) are the interacting nodes, \(t\) is the time of the event and \(k \in \{0, 1\}\) indicates whether the event is a structural evolution, \(k=0\) (edge added) or a communication \(k=1\).

The conditional intensity function \(\lambda\) describes the probability of an event happening. \(\lambda\) is parameterised by two functions \(f\) and \(g\).

\begin{equation}
\lambda_k^{u,v} f_k(g_k^{u,v}(\bar{t}))
\end{equation}

where \(\bar{t}\) is the time just before the current event, \(g\) is a weighted concatenation of node embeddings \(z\), \(g_{k}^{u, v}(\bar{t})=\boldsymbol{\omega}_{k}^{T} \cdot\left[z^{u}(\bar{t}) ; z^{v}(\bar{t})\right]\).
\(f\) is a modified softplus, \(f_{k}(x)=\psi_{k} \log \left(1+\exp \left(x / \psi_{k}\right)\right)\), \(\omega_k\) and \(\psi_k\) are four parameters which enable the temporal point process to be modelled on two different time scales.

The TPP is parameterised by an RNN. The RNN incorporates aggregation of local node embeddings, the previous embedding of the given node and an exogenous drive.
\begin{equation}
\begin{aligned}
z^{v}\left(t_{p}\right)&=\sigma ( 
W^{struct} h_{struct}^{u}(\bar{t}_{p})\\ &\hspace{3.5mm}+W^{rec} z^{v}(\bar{t}_{p}^{v}) W^{t}(t_{p}-\bar{t}_{p}^{v}))
\end{aligned}
\end{equation}

where \(h^u_{s t r u c t}\) is given by an attention mechanism that aggregates embeddings of neighbours of \(u\). The attention mechanism uses an attention matrix \(S\) which is calculated and maintained by the adjacency matrix \(A\) and the intensity function \(\lambda\). In short, the \(\lambda\) parameterises the attention mechanism used by the RNN which in turn is used to parameterise \(\lambda\). Thus \(\lambda\) influences the parameterisation of itself.

With \(\lambda\) well parameterised it serves as a model for the dynamic network and its conditional intensity function can be used to predict link appearance and time of link appearance.

Latent dynamic graph (LDG) \cite{knyazev2019learning} uses Kipf \emph{et al.}'s Neural Relational Inference (NRI) model \cite{kipf2018neural} to extend DyREP. The idea is to re-purpose NRI to encode the interactions on the graph, generate a temporal attention matrix which is then used to improve upon self-attention originally used in DyREP. 

Graph Hawkes Network (GHN) \cite{han2020graph} is another method that parameterizes a TPP through a deep neural architecture. Similarly to Know-Evolve \cite{trivediKnowEvolveDeepTemporal2017}, it targets temporal knowledge networks. A part of the architecture, the Graph Hawkes Process, is an adapted continuous-time LSTM for Hawkes processes \cite{mei2017neural}. 

\subsubsection{Time embedding based models}
Some continuous models rely on time embedding methods. This includes using positional encoding to represent the time dimension as introduced by Vaswani \emph{et al.} \cite{vaswaniAttentionAllYou2017}.  An example of a time embedding method is time2vec \cite{kazemi2019time2vec}. This is a positional encoding, similar to the transformer but especially focused on encoding temporal patterns. Another example, is the functional time embedding introduced by Xu \emph{et al.} \cite{xu_self-attention_2019} which converts learning temporal patterns to the kernel learning problem and learns the kernel function. They apply classical functional analysis to enable functional learning. These time embedding methods are particularly aimed at capturing temporal difference $t_i - t_j$, which is of substantial benefit when modelling interaction networks since it enables them to effectively capture inter-event time.

Temporal Graph Attention (TGAT) \cite{xu2020inductive} was the first continuous DGNN to use a time embedding. The authors use the functional time embedding they introduced separately \cite{xu_self-attention_2019}, however when comparing different versions of the embedding they end up using a non-parametric version (Equation \ref{eq:timeembedding}) which is near identical to time2vec \cite{kazemi2019time2vec}. 

\begin{equation}
\label{eq:timeembedding}
\resizebox{0.89\hsize}{!}{
$
\begin{aligned}
\Phi_d(t, t_1) = &\big[\cos \left(\omega_1\left(t-t_{1}\right) + \varphi_1 \right), \ldots, 
\cos \left(\omega_d\left(t-t_{1}\right) + \varphi_d \right)\big]
\end{aligned}
$
}
\end{equation}

Where $\omega_i$ and $\varphi_i$ are learned weights and $d$ is the size of the time embedding.

A TGAT layer concatenates together the node features, edge features (optional) and time features of each neighbouring node as well as the target node. It then applies masked-attention similar to the attention in GAT \cite{velickovicGraphAttentionNetworks2017}. For each layer added an additional hop of neighbours is added. The authors found 2 layers (2 hops) to be optimal, as additional hops exponentially increase run-time.

\begin{equation}
\label{eq:tgatz}
\begin{aligned}
Z(t)=&\Big[\tilde{h}_{0}^{(l-1)}\left(t\right) \| e_{0,0}\left(t_0\right) \| \Phi_{d_{T}}(0), \\
&\tilde{h}_{1}^{(l-1)}\left(t_{1}\right) \| e_{0,1}\left(t_1\right) \| \Phi_{d_{T}}\left(t-t_{1}\right),\\
&\ldots, \\
&\tilde{h}_{N}^{(l-1)}\left(t_{N}\right) \| e_{0,N}\left(t_N\right) \| \Phi_{d_{T}}\left(t-t_{N}\right)\Big]^{\top}
\end{aligned}
\end{equation}

$Z(t)$ is an entity-temporal feature matrix which include features of nodes, edges and inter-event time. $l$ is the layer. In line with self-attention $Z(t)$ is linearly projected to obtain the 'query', 'key' and 'value'. 

\begin{equation}
\begin{aligned}
&q(t)=[Z(t)]_{0} W_{Q}\\
&K(t)=[Z(t)]_{1: N} W_{K}\\
&V(t)=[Z(t)]_{1: N} W_{V}\\
\end{aligned}
\end{equation}

$[Z(t)]_{0}$ is the features of the target node (the node we want to compute the embedding for) and $[Z(t)]_{1:N}$ is the features of its neighbours. TGAT applies its attention to $Z(t)$ to obtain $h(t)$, the hidden representation of the node. 

\begin{equation}
h(t) = 
\operatorname{softmax}(\frac{q(t)K(t)}{\sqrt{d_k}})V(t)
\end{equation}

Finally, the hidden representation is concatenated with the (static) node embedding of the target node, $x_0$, and passed to a feed-forward network. 

\begin{equation}
\tilde{h}_{0}^{(l)}(t)=\operatorname{FFN}\left(h(t) \| x_{0}\right)
\end{equation}

Temporal Graph Networks (TGN) \cite{rossi2020temporal} extends TGAT by adding a memory module. The memory module embeds the history of the node. The memory vector is added to $Z(t)$ in Equation \ref{eq:tgatz}.

\subsection{Discussion and summary}
\label{sec:org4f5c682}
Deep learning on dynamic graphs is still a new field, however, there are already promising methods that show the capacity to encode dynamic topology. This section has provided a comprehensive and detailed survey of deep learning models for dynamic graph topology encoding.

The encoders are summarised and compared in Table \ref{tab:deepencoders}. Models are listed based on their encoders and the encoders capacity to model link and node dynamics. Any model which cannot model link deletion or link duration can only model strictly evolving networks or interaction networks (see section \ref{sec:dynamicnetworktypes}).  

Table \ref{tab:deepencoders} list many models as not supporting link deletion, it is possible to model link deletion by link deletion events  and thus an interaction network can model a persistent link disappearing. Any continuous model should also be able to model node deletion by removing the node from node neighbourhood aggregation to effectively delete it. However, while these ways of modelling dynamics have been discussed by earlier works \cite{rossi2020temporal}, to the best of our knowledge, they have not been implemented in practice.

Most methods focus on discrete graphs which enable them to leverage recent advances in graph neural networks. This allows for modelling of diverse graphs, including node-dynamic graphs, dynamic labels on nodes and due to the use of snapshots, temporal networks can also be handled. Continuous models currently exist for strictly growing networks and  interaction networks. This leaves many classes of dynamic graphs unexplored. Since continuous models have some inherent advantages over discrete graphs (see section \ref{sec:dynnetsummary}), expanding the repertoire of dynamic network classes for continuous models, is a promising future direction.

All discrete DGNNs use a GNN to model graph topology and a deep time-series model, typically an RNN, to model the time dependency. Two types of architectures can be distinguished: (i) the stacked DGNN and (ii) the integrated DGNN. Different stacked DGNNs only differ in which spatial and temporal layers are used to stack (which GNN they use and which time series layer), while the integrated DGNNs may differ not only by how they model spatial and temporal patterns but also in how they integrate the spatial and temporal modules. Given the same graph, a stacked DGNN would generally have fewer parameters than a typical integrated DGNN (such as GCRN-M2 \cite{seoStructuredSequenceModeling2018}). Both approaches offer great flexibility in terms of which GNN and RNN can be used. They also are rather flexible in that they can model networks with both appearing and disappearing edges as well as dynamic labels.

Discrete models tend to treat every snapshot as a static graph, thus the complexity of the model is proportional to the size of the graph in each snapshot and the number of snapshots. Whereas a continuous model complexity is generally proportional to the number of changes in the graph. If a discrete approach creates snapshots using time-windows, then it can trade off temporal granularity (and thus theoretically modelling accuracy) for faster computation by using larger time-windows for each snapshot.

\begin{table}
\caption{DGNN model types and network types. All continuous DGNNs work on specific types of networks, such as directed or knowledge networks, therefore there are no continuous DGNNs for any general purpose dynamic network.}
\label{tab:deepencodersnetworks}
\centering
\adjustbox{max width=\linewidth}{
\begin{tabular}{c|cccc}

\multirow{2}*{\makecell{DGNN\\ model type}} & \multicolumn{4}{c}{Network type}\\
 & Interaction & Temporal & Evolving & Strictly evolving\\
\hline
Discrete & Yes & Yes & Yes & Yes\\
Continuous & Yes & No & No & No\\
\end{tabular}
}
\end{table}

Table \ref{tab:deepencoders} shows that every continuous DGNN is aimed at a special type of continuous network. This is reflected in Table \ref{tab:deepencodersnetworks} which shows that there is, as of yet, no continuous DGNN encoder for any general-purpose dynamic network. 

\textbf{So which one should you chose?} 
Converting the dynamic network to an edge-weighted network is a simple, and depending on the application, possibly "good enough" approach. A practitioner only need to come up with some scheme to weight edges, and then feed that to an optimized implementation of a standard GNN, e.g. GCN \cite{kipfSemisupervisedClassificationGraph2016} or GAT \cite{velickovicGraphAttentionNetworks2017}. TDGNN \cite{qu_continuous-time_2020} shows a good example of such a scheme by weighting the edges using an exponential distribution, which weights more recent edges higher than old edges.

Another approach which should be considered before trying any large DGNN model is whether applying a static GNN on a discrete representation might yield good enough results. Given the same number of features and layer size, it will train faster and generally be a simpler model.

The choice between discrete and continuous depends on the data and the intended problem. If temporal granularity and performance is not a concern then one of the advanced discrete approaches such as DySAT or EvolveGCN will likely be a great fit for most dynamic network problems. Since they naturally support link deletion, node addition and node deletion, they provide good general-purpose functionality. 

The Discrete DGNNs covered in this work all iterate over snapshots to encode, while the continuous DGNNs iterate edge-by-edge. The continuous therefore tend to take longer to train compared to the discrete models. This is especially true if the network is rather dense.

Evolving networks are well served by any discrete approach, however, with the recent dominance of attention architectures \cite{vaswaniAttentionAllYou2017}, we would expect DySAT to do well in a comparative test. EvolveGCN is expected to train fast on an evolving network with little change between snapshots. The discrete methods are also suited for temporal networks given that the length of the time-windows covered by snapshots is well selected.

If node dynamics is an important feature of the network you wish to model, then you should choose a model that can encode node dynamics such as DySAT \cite{sankarDynamicGraphRepresentation2018}, EvolveGCN \cite{parejaEvolveGCNEvolvingGraph2019} or HDGNN \cite{zhou2020heterogeneous}. 

 If you have an interaction network with detailed timestamps, then TGAT \cite{xu2020inductive} or TGN \cite{rossi2020temporal} are likely good fits. If run-time complexity and time granularity are essential to the dynamic complex network at hand (for example in the case of a temporal network), then non-deep learning methods that are not covered by this survey are recommended. Those methods can be explored in the literature referred to in section \ref{sec:networkmodels}.


\begin{table*}
\caption{Deep encoders for dynamic network topology. While we note which GNNs are used in each of the discrete models it is usually trivial to replace it with another GNN.}
\label{tab:deepencoders}
\adjustbox{max width=\linewidth}{
\begin{tabular}{llllllll}
Model type & Model name & Encoder & Link addition & Link deletion & Node addition & Node deletion & Network type\\
\hline
Discrete networks & & & & & & &\\
Stacked DGNN & GCRN-M1  \cite{seoStructuredSequenceModeling2018} & Spectral GCN \cite{defferrardConvolutionalNeuralNetworks2016} \& LSTM & Yes & Yes & No & No & Any\\
& WD-GCN \cite{manessiDynamicGraphConvolutional2020} & Spectral GCN \cite{kipfSemisupervisedClassificationGraph2016} \& LSTM & Yes & Yes & No &No & Any\\
& CD-GCN \cite{manessiDynamicGraphConvolutional2020} & Spectral GCN \cite{kipfSemisupervisedClassificationGraph2016} \& LSTM & Yes & Yes & No & No & Any\\
& RgCNN \cite{narayanLearningGraphDynamics2018} & Spatial GCN \cite{niepertLearningConvolutionalNeural2016} \& LSTM & Yes & Yes & No & No & Any\\
& DyGGNN \cite{taheriLearningRepresentEvolution2019} & GGNN \cite{liGatedGraphSequence2017} \& LSTM & Yes & Yes & No & No & Any\\
& DySAT \cite{sankarDynamicGraphRepresentation2018} & GAT \cite{velickovicGraphAttentionNetworks2017} \& temporal attention \cite{vaswaniAttentionAllYou2017} & Yes & Yes & Yes & Yes & Any\\
& TNDCN \cite{wang2020generic,wang2020tedic} & Spectral GCN \cite{wang2020generic} \& TCN \cite{wang2020generic} & Yes & Yes & No & No & Any\\
& StrGNN \cite{cai2020structural} & Spectral GCN \cite{kipfSemisupervisedClassificationGraph2016} \& GRU & Yes & Yes & No & No & Any\\
& HDGNN \cite{zhou2020heterogeneous} & Spectral GCN \cite{kipfSemisupervisedClassificationGraph2016} \& A variety of RNNs & Yes & Yes & Yes & Yes & Heterogeneous\\
& TeMP \cite{wu2020temp} & R-GCN \cite{schlichtkrullModelingRelationalData2018} stacked with either GRU or attention & Yes & Yes & No & No & Knowledge\\
Integrated DGNN & GCRN-M2 \cite{seoStructuredSequenceModeling2018} & GCN \cite{defferrardConvolutionalNeuralNetworks2016} integrated in an LSTM & Yes & Yes & No & No & Any\\
& GC-LSTM \cite{chenGCLSTMGraphConvolution2018} & GCN \cite{defferrardConvolutionalNeuralNetworks2016} integrated in an LSTM & Yes & Yes & No & No & Any\\
& EvolveGCN \cite{parejaEvolveGCNEvolvingGraph2019} & LSTM integrated in a GCN \cite{kipfSemisupervisedClassificationGraph2016} & Yes & Yes & Yes & Yes & Any\\
& LRGCN \cite{liPredictingPathFailure2019} & R-GCN \cite{schlichtkrullModelingRelationalData2018} integrated in an LSTM & Yes & Yes & No & No & Any\\
& RE-Net \cite{jin2019recurrent} & R-GCN \cite{schlichtkrullModelingRelationalData2018} integrated in several RNNs & Yes & Yes & No & No & Knowledge\\
& TNA \cite{bonner2019temporal} & GCN \cite{kipfSemisupervisedClassificationGraph2016} stacked with a GRU and a linear layer & Yes & Yes & No & No & Any\\
\hline
Continuous networks & & & & & & &\\
RNN based & & & & & & &\\
& Streaming GNN \cite{maStreamingGraphNeural2018} & \makecell{Node embeddings maintained by \\ architecture consisting of T-LSTM \cite{baytasPatientSubtypingTimeAware2017}} & Yes & No & Yes & No & Directed strictly evolving\\
& JODIE \cite{kumar2019predicting} & \makecell{Node embeddings maintained by\\ an RNN based architecture} & Yes & No & No & No & Bipartite, interaction\\
TTP based & & & & & & &\\
& Know-Evolve \cite{trivediKnowEvolveDeepTemporal2017} & \makecell{TPP parameterised by an RNN} & Yes & No & No & No & Interaction, knowledge network\\
& DyREP \cite{trivedi2018dyrep} & \makecell{TPP parameterised by an RNN\\ aided by structural attention} & Yes & No & Yes & No & Interaction combined with strictly evolving\\
& LDG \cite{knyazev2019learning} & \makecell{TPP, RNN and self-attention} & Yes & No & Yes & No & Interaction combined with strictly evolving\\
& GHN \cite{han2020graph} & \makecell{TPP parameterised by a\\ continuous time LSTM \cite{mei2017neural}} & Yes & No & No & No & Interaction, knowledge network\\
 Time embedding based& & & & & & &\\
& TGAT \cite{xu2020inductive} & \makecell{Temporal \cite{kazemi2019time2vec} and structural \cite{velickovicGraphAttentionNetworks2017}\\ attention} & Yes & No & Yes & No & Directed or undirected interaction\\
& TGN \cite{rossi2020temporal} & \makecell{Temporal \cite{kazemi2019time2vec} and structural \cite{velickovicGraphAttentionNetworks2017}\\ attention with memory} & Yes & No & Yes & No & Directed or undirected interaction\\
\hline
\end{tabular}
}
\end{table*}

\section{Deep learning for prediction of network topology}
\label{sec:orge8000db}
\label{sec:dynamiclinkprediction}
Any embedding method can be thought of as a concatenation of an encoder and a decoder \cite{hamiltonRepresentationLearningGraphs2017}. Until now, we have discussed encoders, but the quality of embeddings depend on the decoder and the loss function as well. While the encoders in Section \ref{sec:deeplearningmodels} can be paired with a variety of decoders and loss functions depending on the intended task, we focus in this section on one of the most commonly tackled problems - link prediction.

Prediction problems can be defined for many different contexts and settings. In this survey, we refer to the prediction of the future change to the network topology. Much work has been done on the prediction of missing links in networks, which can be thought of as an interpolation task. This section explores how dynamic graph neural networks can be used for link prediction and deal exclusively with the extrapolation (future link prediction) task. 

Predictions can be done in a time-conditioned or time-predicting manner \cite{kazemiRelationalRepresentationLearning2019}. Time-predicting means that a method predicts \emph{when} an event will occur and time-conditioned means that a method predicts \emph{whether} an event will occur at a given time \(t\). For example, if the method predicts the existence of a link in the next snapshot, it is a time-conditioned prediction. If it predicts when a new link between nodes will appear, it is a time-predicting prediction.

Prediction of links often focuses only on the prediction of the appearance of a link. However, link disappearance is less explored but also important for the prediction of network topology. We refer to link prediction based on a dynamic network as dynamic link prediction.

For embedding methods, what is predicted and how is decided by the decoder. You can have both time-predicting and time-conditioned decoders. The prediction capabilities will depend on the information captured by the embeddings. Thus, an embedding that captures continuous-time information has a higher potential to model temporal patterns. Well modelled temporal and structural embeddings offer a better foundation for a decoder and thus potentially better predictions.

If dealing with discrete data and few timestamps, a time-conditioned decoder can be used for time prediction. This can be done by applying the time-conditioned decoder to every candidate timestamp \(t\) and then consider the \(t\) where the link has the highest probability of appearing.

The rest of this section is a description of how the surveyed models from the previous section can be used to perform predictions. This includes mainly a discussion on decoders and loss functions. Since the surveyed models aim to predict the time-conditioned existence of links, the focus will be on the dynamic link prediction task.

Autoencoders can use the same decoders and loss functions as other methods. Their aim is typically a little different. The decoder is targeted at the already observed network and tries to recreate the snapshot. A prediction for a snapshot at time \(t+1\) is marginally different from the decoder of an autoencoder which is targeted at already observed snapshots. 

\subsection{Decoders}
\label{sec:org35ddd5c}
Of the surveyed approaches which apply a predicting decoder, almost all  apply a time-conditioned decoder. A prediction is then often an adjacency matrix \(\hat{A}^{\tau}\) which indicates the probabilities of an edge at time \(\tau\). Often \(\tau = t+1\).

We consider decoders to be the part of the architecture that produces \(\hat{A}^{\tau}\) from \(Z\) the dynamic graph embeddings.

Since encoders make node embeddings and predicting a link involves two nodes decoders tend to aggregate two node embeddings to predict a link. The simplest way to aggregate is to apply an operator, e.g. the inner product \cite{kipfVariationalGraphAutoencoders2016} (shown in Equation \ref{eq:innerproductdecoder}), concatenation, mean or Hadamard product \cite{qu_continuous-time_2020}. This combines the node embeddings and gives a probability of a link appearing. These simple approaches require that the encoder is able to embed the nodes in a space such that nodes that are likely to connect are close to each other or otherwise able to be decoded by the simple decoder.

Another simple decoder is to use a simple feed-forward network. The decoder as before receives two node embeddings and gives out a probability for whether the link appeared or didn't appear. This approach is used by several models for link prediction \cite{parejaEvolveGCNEvolvingGraph2019,goyalDyngraph2vecCapturingNetwork2019,chenELSTMDDeepLearning2019}. While this requires more parameters, the decoder is can easily be dwarfed in size by the encoder and it enables decoding of non-linear relationships between node embeddings. 

\begin{equation}
\label{eq:innerproductdecoder}
p\left(A^t_{i j}=1 | z^{t}_{i}, z^{t}_{j}\right)=\sigma\left((z^{t}_{i})^{\top} z^{t}_{j}\right)
\end{equation}

Where \(z_k\) is the node embedding of node \(k\). An inner product decoder works well if we only want to predict or reproduce the graph topology. If we would like to decode the feature matrix then a neural network should be used \cite{hajiramezanaliVariationalGraphRecurrent2019}.

Wu \emph{et al.} \cite{wu2020evonet} uses GraphRNN, a deep sequential generative model as a decoder \cite{youGraphRNNDeepGenerative2018}. What is unique with GraphRNN is that it reframes the graph generation problem as a sequential problem. The GraphRNN authors claim increased performance over feed-forward auto-encoders. 

In general, there are many options for how decoding can be done. A decoder might be viable as long as the probability for each edge is produced from the latent variables and the architecture can be efficiently optimized with back-propagation.

The only surveyed method using a time-predicting decoder is DyRep \cite{trivedi2018dyrep}. DyRep uses the conditional intensity function of its temporal point process to model the dynamic network. 

While the focus in this section is on decoders that are used directly for the forecasting task, it is important to note that downstream learning can also be used. This is the DGNN trained on a task and the node embeddings are used for a different task. For example, the DGNN can be trained on node classification and then the node embeddings are used later for link prediction. An example of this is seen in \cite{sankarDynamicGraphRepresentation2018}, where a logistic regression classifier is trained on the node embeddings of snapshot \(t\) to predict links at \(t+1\).

\subsection{Loss functions}
\label{sec:org3f5d03c}
The loss function is central to any deep learning method, as it is the equation that is being optimized. Regarding loss functions, we can make a distinction between (i) link prediction optimizing methods; and (ii) autoencoder methods. As the prediction methods optimize towards link prediction directly, an autoencoder optimizes towards the recreation of the dynamic graph. Despite have slightly different aims, both approaches have been used for link prediction and have been shown to perform well.

\subsubsection{Link prediction}
\label{sec:orgbafed97}
Prediction of edges is seen as a binary classification task. Traditional link prediction is well known for being extremely unbalanced \cite{yangEvaluatingLinkPrediction2015,junuthulaEvaluatingLinkPrediction2016}. For predicting methods the loss function is often simply the binary cross-entropy \cite{taheriLearningRepresentEvolution2019,parejaEvolveGCNEvolvingGraph2019,sankarDynamicGraphRepresentation2018}.

Some models use negative sampling \cite{parejaEvolveGCNEvolvingGraph2019,sankarDynamicGraphRepresentation2018}. This transforms the problem of link prediction from a multiple output classification (a prediction for each link) to a binary classification problem (is the link a "good" link or a "bad" link). This speeds up computation and deals with the well-known class imbalance problem in link prediction. The rate of negative samples used vary from work to work, EvolveGCN \cite{parejaEvolveGCNEvolvingGraph2019} use 1 to 100 for training, while TGAT \cite{xu2020inductive} and TGN \cite{rossi2020temporal} use 1 to 1. 

\begin{equation}
\label{eqn:lossce}
\mathcal{L}_{C E}=\sum_{i=1}^{n} \sum_{j=1}^{n} A_{i j}^{t} \log(\hat{A}^{t}_{i j})
\end{equation}

Equation \ref{eqn:lossce} is an example of a binary cross entropy loss adapted from \cite{chenGCLSTMGraphConvolution2018}. 

DySAT \cite{sankarDynamicGraphRepresentation2018} sums the loss function only over nodes that are in the same neighbourhood at time \(t\). The neighbourhoods are extracted by taking nodes that co-occur in random walks on the graph. The inner product is calculated as a part of the summation in the loss function. This means that the inner product will be calculated only for the node pairs that the loss is computed on. Together it reduces the number of nodes that are summed up and should result in a training speed up. Any accuracy trade-off is not discussed by the authors.

\subsubsection{Autoencoders}
\label{sec:org2012ec7}
Autoencoder approaches \cite{goyal2018dynamicgem,goyalDyngraph2vecCapturingNetwork2019,chenELSTMDDeepLearning2019} aim to reconstruct the dynamic network. All surveyed autoencoders operate on discrete networks. Therefore the reconstruction of the network is reduced to the reconstruction of each snapshot. This entails creating a loss function that penalizes wrong reconstruction of the input graph. Variational autoencoder approaches \cite{xuGenerativeGraphConvolutional2019,hajiramezanaliVariationalGraphRecurrent2019} also aim to be generative models. To be generative, they need to enable interpolation in latent space. This is achieved by adding a term to the loss function which penalizes the learned latent variable distribution for being different from a normal distribution. It is also common to add regularization to the loss functions to avoid overfitting.

\begin{equation}
\label{eqn:losselstmd}
\mathcal{L}=\sum_{i=1}^{n} \sum_{j=1}^{n}\left(A^t_{ij}-\hat{A}^t_{ij}\right) * P_{ij}
\end{equation}

Equation \ref{eqn:losselstmd} is the reconstruction penalizing component of E-LSTM-D's loss function \cite{chenELSTMDDeepLearning2019}. \(P\) is a matrix which increases the focus on existing links. \(p_{ij} = 1\) if \(A^t_{ij} = 0\) and \(p_{ij} = \beta > 1\) if \(A^t_{ij} = 1\).

\subsubsection{Temporal Point Processes}
DyRep \cite{trivedi2018dyrep} models a dynamic network by parameterising a temporal point process. Its loss function influences how the temporal point process is optimized.

\begin{equation}
\mathcal{L}=-\sum_{p=1}^{P} \log \left(\lambda_{p}(t)\right)+\int_{0}^{T} \Lambda(\tau) d \tau
\end{equation}

where $P$ is the set of observed events, $\lambda$ is the intensity function and  $\Lambda(\tau)=\sum_{u=1}^{n} \sum_{v=1}^{n} \sum_{k \in\{0,1\}} \lambda_{k}^{u, v}(\tau)$
 is the survival probability for all events that did not happen. Survival probability indicates the probability of an event not happening \cite{aalen2008survival}. The first term thus rewards a high intensity when an event happens, whereas the second term rewards a low intensity (high survival) of events that do not happen. 
 
 Trivedi \emph{et al.} \cite{trivedi2018dyrep} further identify that calculating the integral of $\Lambda$ is intractable. They get around that by sampling non-events and estimating the integral using Monte Carlo estimation, this is done for each mini-batch.
 
\subsubsection{Regularization}
There are several different approaches for adding regularization to loss functions to avoid overfitting. The total loss function (equation \ref{eqn:losstotal}) is composed of the reconstruction loss and the regularization with an optional constant \(\alpha\) to balance the terms. Here we cover the methods that use regularization, however many models chose to not use regularization as they find that they don't have a problem with overfitting \cite{parejaEvolveGCNEvolvingGraph2019,trivedi2018dyrep,xu2020inductive,rossi2020temporal}.

\begin{equation}
\label{eqn:losstotal}
\mathcal{L}_{\text{total}} = \mathcal{L} + \alpha\mathcal{L}_{reg}
\end{equation}

A common way to regularize is through summing up all the weights of the model, thus keeping the weights small and the model less likely to overfit. The \(L_2\) norm is commonly used for this \cite{chenGCLSTMGraphConvolution2018,chenELSTMDDeepLearning2019}.

The variational autoencoder methods use a different regularizer. They normalize the node embeddings compared to a prior. In traditional variational autoencoders, this prior is a Normal distribution with mean \(0\) and standard deviation \(1\). In dynamic graph autoencoders \cite{hajiramezanaliVariationalGraphRecurrent2019,xuGenerativeGraphConvolutional2019}, the prior is still a Gaussian, but it is parameterised by previous observations. Equation \ref{eqn:kldgnn} is the regularization term from \cite{hajiramezanaliVariationalGraphRecurrent2019}.

\begin{equation}
\label{eqn:kldgnn}
\begin{aligned}
\mathbf{K L}(&q\left(Z^{t} | A^{\leq t}, X^{\leq t}, Z^{<t}\right) \\
&\| p\left(Z^{t} | A^{<t}, X^{<t}, Z^{<t}\right))
\end{aligned}
\end{equation}

where \(q\) is the encoder distribution and \(p\) is the prior distribution. \(\mathbf{KL}\) is the Kullback-Leibler divergence which measures the difference between two distributions. The \(A^{<t}\) indicate all adjacency matrices up to, but not including \(t\) and similarly for the other matrices. We can see that the prior is influenced by previous snapshots, but not by the current. Whereas the encoder is influenced by the previous and the current snapshot.

\subsection{Evaluation metrics}
\label{sec:orgefa15b3}
Link prediction is plagued by high class imbalance. It is a binary classification, a link either exists or not and most links will not exist. In fact, actual links tend to constitute less than 1\% of all possible links \cite{liDeepDynamicNetwork2018}. AUC and precision@k are two commonly used evaluation metrics in static link prediction \cite{yangEvaluatingLinkPrediction2015,luLinkPredictionComplex2011}. If dynamic link prediction requires the prediction of both appearing and disappearing edges, the evaluation metric needs to reflect that. Furthermore, traditional link prediction metrics have shortcomings when used in a dynamic setting \cite{junuthulaEvaluatingLinkPrediction2016}.

For a detailed discussion on the evaluation of link prediction, we refer to Yang \emph{et al.} \cite{yangEvaluatingLinkPrediction2015} for static link prediction and Junuthula \emph{et al.} \cite{junuthulaEvaluatingLinkPrediction2016} for dynamic link prediction evaluation.

\begin{enumerate}
\item \textbf{Area under the curve (AUC).}
\label{sec:orgfd678e3}
The area under the curve (AUC) is used to evaluate a binary classification and has the advantage of being independent of the classification threshold. The AUC is the area under the receiver operating characteristic (ROC) curve. The ROC is a plot of the true positive rate and the false positive rate.

The AUC evaluates predictions based on how well the classifier ranks the predictions, this provides a measure that is invariant of the classification threshold. In link prediction, there has been little research into finding the optimal threshold \cite{martinezSurveyLinkPrediction2016}, using the AUC for evaluation avoids this problem.

Yang \emph{et al.} \cite{yangEvaluatingLinkPrediction2015} note that AUC can show deceptively high performance in link prediction due to the extreme class imbalance. They recommend the use of PRAUC instead.

\item \textbf{PRAUC.}
\label{sec:org1375674}
The PRAUC is similar to the AUC except that it is the area under the precision-recall curve. The metric is often used in highly imbalanced information retrieval problems \cite{junuthulaEvaluatingLinkPrediction2016}.

PRAUC is recommended by Yang \emph{et al.} \cite{yangEvaluatingLinkPrediction2015} as a suitable metric for traditional (static) link prediction due to the deceptive nature of the ROC curve and because PRAUC shows a more discriminative view of classification performance. And recommended for the same reasons by Li \emph{et al.} for dynamic link prediction \cite{liDeepDynamicNetwork2018}.

One way of calculating the PRAUC is by using Mean Average Precision (MAP). MAP is the mean of the average precision (AP) per node.

\item \textbf{Fixed-threshold metrics.}
\label{sec:orgd5b5bd4}
One of the most common fixed threshold metrics in traditional link prediction is Precision@k. It is the ratio of items that are correctly predicted. From the ranking prediction the top \(k\) predictions are selected, then precision is the ratio \(\frac{k_r}{k}\), where \(k_r\) is the number of correctly predicted links in the top \(k\) predictions.

While a higher precision indicates a higher prediction accuracy, it is dependent on the parameter \(k\). \(k\) might be given on web-scale information retrieval, where we care about the accuracy of the highest \(k\) ranked articles, in link prediction it is difficult to find the right cut-off \cite{martinezSurveyLinkPrediction2016}.

A fixed-threshold can be applied to other common metrics including accuracy, recall and F1 among others \cite{yangEvaluatingLinkPrediction2015}. These methods suffer from instability in their predictions, where a change of thresholds can lead to contradictory results \cite{yangEvaluatingLinkPrediction2015}. This problem is also observed in dynamic link prediction \cite{junuthulaEvaluatingLinkPrediction2016}. Fixed-threshold metrics are not recommended unless the targeted problem has a natural threshold \cite{yangEvaluatingLinkPrediction2015}.

\item \textbf{Sum of absolute differences (SumD).}
\label{sec:org726efaa}
Li \emph{et al.} \cite{liDeepDynamicNetwork2018} pointed out that models often have similar AUC scores and suggested SumD as a stricter measurement of accuracy. It is simply, the number of mispredicted links. The metric has different meanings depending on how many values are predicted since it is not normalized according to the total number of links. Chen \emph{et al.} considers SumD misleading for this reason \cite{chenELSTMDDeepLearning2019}. The metric strictly punishes false positives, since there are so many links not appearing, a slightly higher rate of false positives will have a large impact on this metric.

\item \textbf{Error rate.}
\label{sec:org669a99b}
Since SumD suffers from several drawbacks an extension is suggested by Chen \emph{et al.} \cite{chenELSTMDDeepLearning2019}. Error rate normalizes SumD by the total number of existing links. 

\begin{equation}
\text { Error Rate }=\frac{N_{\text {false }}}{N_{\text {true }}}
\end{equation}

where \(N_{\text {false}}\) is the number of mispredicted links and \(N_{\text {true}}\) is the number of existing links. The error rate is very similar to recall, except that recall focuses on true positives, where the error rate focuses on false positives. Another difference between recall and error rate is that recall is normalized between 0 and 1, while the error rate may be above 1 if the number of mispredicted links outnumber the number of existing links The error rate is a good metric if the number of false positives is a major concern. In dynamic link prediction, false positives become a major issue due to the massive class imbalance of the prediction problem. 
 
\item \textbf{GMAUC.}
\label{sec:org681948b}
After a thorough investigation of evaluation metrics for dynamic link prediction, Junuthula \emph{et al.} suggests GMAUC as an improvement over other metrics \cite{junuthulaEvaluatingLinkPrediction2016}. The key insight is that dynamic link prediction can be divided into two sub-problems: (i) predicting the disappearance of links that already exist or the appearance of links that have once existed; and (ii) predicting links that have never been seen before. When the problem is divided in this way, each of the sub-problems takes on different characteristics.

Prediction of links that have never been seen before is equivalent to traditional link prediction, for which PRAUC is a suitable metric \cite{yangEvaluatingLinkPrediction2015}.  Prediction of already existing links is both the prediction of once seen links appearing and existing links disappearing. This is a more balanced problem than traditional link prediction, thus AUC is a suitable measure. \cite{junuthulaEvaluatingLinkPrediction2016} note that both the mean and the harmonic mean will lead to either the AUC or the PRAUC to dominate, thus the geometric mean is used to form a unified metric.

\begin{equation}
\resizebox{.7\hsize}{!}{
$
\begin{aligned}
&\mathrm{GMAUC}=\\
&\hspace{7mm}\sqrt{\frac{\mathrm{PRAUC}_{\mathrm{new}}-\frac{P}{P+N}}{1-\frac{P}{P+N}} \cdot 2\left(\mathrm{AUC}_{\mathrm{prev}}-0.5\right)}
\end{aligned}
$
}
\end{equation}

\(PRAUC_{new}\) is the PRAUC score of new links, \(AUC_{prev}\) is the AUC score of previously observed links.

The authors note the advantages of GMAUC:
\begin{itemize}
\item Based on threshold curves, thus avoids the pitfall of fixed-threshold metrics
\item Accounts for differences between predicting new and already observed edges without having the metric to be dominated by either sub-problem.
\item Any predictor that predicts only new edges or only previously observed edges gets a score of 0.
\end{itemize}
\end{enumerate}

However, it does hinge on the assumption that reoccurring edges is a balanced enough prediction problem that AUC is suitable. And that is not necessarily the case. Many real-world networks are much more sparse than the two networks used by Junuthula \emph{et al.} \cite{junuthulaEvaluatingLinkPrediction2016}.

\subsection{Discussion and summary}
\label{sec:org9f50c37}
In this section we have provided an overview of how, given a dynamic network encoder, one can perform network topology prediction. The overview includes how methods from section \ref{sec:deeplearningmodels} use their embeddings for prediction. This completes the journey from establishing a dynamic network, to encoding the dynamic topology, to predicting changes in the topology.

Prediction using a deep model requires decoding and the use of a loss function that captures temporal and structural information. Prediction is largely focused on time-conditioned link prediction and the two main modelling approaches are (1) an architecture directly aimed at prediction; and (2) an architecture aimed at generating node embeddings which are then used for link prediction in a downstream step. Most dynamic network models surveyed fall into the second category, including all autoencoder approaches. All else being equal we would expect an architecture directly aimed at prediction to perform better than a two step architecture. This is because the first case will allow the entire architecture to optimize itself towards the prediction task. 

The massive class imbalance makes the evaluation of dynamic link prediction is non-trivial. If the target problem has a natural fixed threshold, then adding a fixed threshold to a common metric such as F1 is likely a good fit. PRAUC (MAP) and Error rate are good metrics that avoids the class imbalance problem and are suitable for both link prediction and dynamic link prediction. The GMAUC metric incorporates the observation that reappearing and disappearing links are not an imbalanced classification. Usage of GMAUC however hinges on the assumption that reoccurring links are a reasonably balanced classification, this is not necessarily true and depends on the data. An evaluation of new methods should report the PRAUC of newly appearing links and the PRAUC or AUC of reappearing links separately. The combined score should also be reported as either the PRAUC, Error rate or GMAUC.

Prediction on dynamic networks is in its infancy. Deep models are largely focused on unattributed time-conditioned discrete link appearance prediction.  This leaves opportunities for future work in a large range of prediction tasks, with some types of prediction still unexplored. Prediction based on continuous-time encoders is a particularly interesting frontier due to the representations inherent advantages and due to the limited amount of works in that area.

\section{Challenges and Future work}
There are plenty of challenges and multiple avenues for the improvement of deep learning for both modelling and prediction of network topology. 

\textbf{Expanding modelling and prediction repertoire.} In this work we have exclusively focused on dynamic network topology. However, complex networks are diverse and not only topology may vary. Topology dynamics can be represented as a 3-dimensional cube (Section \ref{sec:dynnetframework}). However, real networks can be much more complex. Complex networks may have dynamic node and edge attributes, they may have directed and/or signed edges, be heterogeneous in terms of nodes and edges and be multilayered or multiplex. Each of these cases can be considered another dimension in the dynamic network hypercube. Designing deep learning models for encoding these network cases expand the repertoire of tasks on which deep learning can be applied. Which types of networks can be encoded can be expanded as well as an expansion of what kind of predictions can be made on those networks. For example, most DGNN models (and most GNN models) encode attributed dynamic networks but predict only graph topology without the node attributes. 

\textbf{Adoption of advances in closely related fields.} Dynamic graph neural networks are based on GNNs and thus advances made to GNNs trickle down and can improve DGNNs. Challenges for GNNs include increasing modelling depth as GNNs struggle with vanishing gradients \cite{DeepGCNsCanGCNs} and increasing scalability for large graphs \cite{hamiltonInductiveRepresentationLearning2017}. As advancements are made in deep neural networks for time series and in GNNs these advancements can be applied to dynamic network modelling and prediction to improve performance. Similarly, improvements in deep time-series modelling can easily be adapted to improve DGNNs.

\textbf{Continuous DGNNs.} Modelling temporal patterns is what distinguishes modelling dynamic graphs from modelling static graphs. Capturing these temporal patterns is key to making accurate predictions. However, most models rely on snapshots which are coarse-grained temporal representations. Methods modelling network change in continuous time will offer fine-grained temporal modelling. Future work is needed for modelling and prediction of continuous-time dynamic networks. 

\textbf{Scalability.} Large scale datasets is a challenge for dynamic network modelling. Real-world datasets tend to be so large that modelling becomes prohibitively slow. Dynamic networks either use a discrete representation in the form of snapshots, in which case processing of each snapshot is the bottleneck or continuous-time modelling which scales with the number of interactions. A snapshot model will need to have frequent snapshots in order to achieve high temporal granularity. In addition, frequent snapshots might undermine the capacity to model a temporal network. Improvements in continuous-time modelling are likely to improve the performance of deep learning modelling on dynamic networks both in terms of temporal modelling capacity and ability to handle large networks.

Dynamic graph neural networks is a new exciting research direction with a broad area of applications.  With these opportunities, the field is ripe with potential for future work.  
\bibliographystyle{unsrt}
\bibliography{refs}

\begin{thebibliography}{100}

\bibitem{strogatzExploringComplexNetworks2001}
Steven~H. Strogatz.
\newblock Exploring complex networks.
\newblock {\em Nature}, 410(6825):268, March 2001.

\bibitem{holmeTemporalNetworks2012}
Petter Holme and Jari Saram{\"a}ki.
\newblock Temporal networks.
\newblock {\em Physics reports}, 519(3):97--125, 2012.

\bibitem{aggarwalEvolutionaryNetworkAnalysis2014}
Charu Aggarwal and Karthik Subbian.
\newblock Evolutionary network analysis: {{A}} survey.
\newblock {\em ACM Computing Surveys (CSUR)}, 47(1):10, 2014.

\bibitem{liRestrictedBoltzmannMachineBased2018}
Taisong Li, Bing Wang, Yasong Jiang, Yan Zhang, and Yonghong Yan.
\newblock Restricted {{Boltzmann Machine}}-{{Based Approaches}} for {{Link
  Prediction}} in {{Dynamic Networks}}.
\newblock {\em IEEE Access}, 6:29940--29951, 2018.

\bibitem{michailElementsTheoryDynamic2018}
Othon Michail and Paul~G Spirakis.
\newblock Elements of the theory of dynamic networks.
\newblock {\em Communications of the ACM}, 61(2):72--81, 2018.

\bibitem{holmeModernTemporalNetwork2015}
Petter Holme.
\newblock Modern temporal network theory: {{A}} colloquium.
\newblock {\em The European Physical Journal B}, 88(9):234, September 2015.

\bibitem{carleyInteroperableDynamicNetwork2007}
Kathleen~M Carley, Jana Diesner, Jeffrey Reminga, and Maksim Tsvetovat.
\newblock Toward an interoperable dynamic network analysis toolkit.
\newblock {\em Decision Support Systems}, 43(4):1324--1347, 2007.

\bibitem{zhouGraphNeuralNetworks2018}
Jie Zhou, Ganqu Cui, Zhengyan Zhang, Cheng Yang, Zhiyuan Liu, Lifeng Wang,
  Changcheng Li, and Maosong Sun.
\newblock Graph {{Neural Networks}}: {{A Review}} of {{Methods}} and
  {{Applications}}.
\newblock {\em arXiv:1812.08434 [cs, stat]}, December 2018.

\bibitem{masudaGuideTemporalNetworks2016}
Naoki Masuda and Renaud Lambiotte.
\newblock {\em A Guide to Temporal Networks}.
\newblock {World Scientific}, 2016.

\bibitem{casteigtsTimeVaryingGraphsDynamic2011}
Arnaud Casteigts, Paola Flocchini, Walter Quattrociocchi, and Nicola Santoro.
\newblock Time-{{Varying Graphs}} and {{Dynamic Networks}}.
\newblock In Hannes Frey, Xu~Li, and Stefan Ruehrup, editors, {\em Ad-Hoc,
  {{Mobile}}, and {{Wireless Networks}}}, pages 346--359, {Berlin, Heidelberg},
  2011. {Springer Berlin Heidelberg}.

\bibitem{kazemiRelationalRepresentationLearning2019}
Seyed~Mehran Kazemi, Rishab Goel, Kshitij Jain, Ivan Kobyzev, Akshay Sethi,
  Peter Forsyth, and Pascal Poupart.
\newblock Relational {{Representation Learning}} for {{Dynamic}}
  ({{Knowledge}}) {{Graphs}}: {{A Survey}}.
\newblock {\em arXiv:1905.11485 [cs, stat]}, May 2019.

\bibitem{xie2020survey}
Yu~Xie, Chunyi Li, Bin Yu, Chen Zhang, and Zhouhua Tang.
\newblock A survey on dynamic network embedding.
\newblock {\em arXiv preprint arXiv:2006.08093}, 2020.

\bibitem{barros2021survey}
Claudio~DT Barros, Matheus~RF Mendonça, Alex~B Vieira, and Artur Ziviani.
\newblock A survey on embedding dynamic graphs.
\newblock {\em arXiv preprint arXiv:2101.01229}, 2021.

\bibitem{jain2016structural}
Ashesh Jain, Amir~R Zamir, Silvio Savarese, and Ashutosh Saxena.
\newblock Structural-rnn: Deep learning on spatio-temporal graphs.
\newblock In {\em Proceedings of the ieee conference on computer vision and
  pattern recognition}, pages 5308--5317, 2016.

\bibitem{yu2017spatio}
Bing Yu, Haoteng Yin, and Zhanxing Zhu.
\newblock Spatio-temporal graph convolutional networks: A deep learning
  framework for traffic forecasting.
\newblock {\em arXiv preprint arXiv:1709.04875}, 2017.

\bibitem{parejaEvolveGCNEvolvingGraph2019}
Aldo Pareja, Giacomo Domeniconi, Jie Chen, Tengfei Ma, Toyotaro Suzumura,
  Hiroki Kanezashi, Tim Kaler, and Charles~E. Leisersen.
\newblock Evolvegcn: Evolving graph convolutional networks for dynamic graphs.
\newblock {\em CoRR}, abs/1902.10191, 2019.

\bibitem{sankarDynamicGraphRepresentation2018}
Aravind Sankar, Yanhong Wu, Liang Gou, Wei Zhang, and Hao Yang.
\newblock Dysat: Deep neural representation learning on dynamic graphs via
  self-attention networks.
\newblock In James Caverlee, Xia~(Ben) Hu, Mounia Lalmas, and Wei Wang,
  editors, {\em {WSDM} '20: The Thirteenth {ACM} International Conference on
  Web Search and Data Mining, Houston, TX, USA, February 3-7, 2020}, pages
  519--527. {ACM}, 2020.

\bibitem{xu2020inductive}
Da~Xu, Chuanwei Ruan, Evren Korpeoglu, Sushant Kumar, and Kannan Achan.
\newblock Inductive representation learning on temporal graphs.
\newblock {\em arXiv preprint arXiv:2002.07962}, 2020.

\bibitem{rossi2020temporal}
Emanuele Rossi, Ben Chamberlain, Fabrizio Frasca, Davide Eynard, Federico
  Monti, and Michael Bronstein.
\newblock Temporal graph networks for deep learning on dynamic graphs.
\newblock {\em arXiv preprint arXiv:2006.10637}, 2020.
\newblock tex.ids= rossiTemporalGraphNetworks2020.

\bibitem{chenGCLSTMGraphConvolution2018}
Jinyin Chen, Xuanheng Xu, Yangyang Wu, and Haibin Zheng.
\newblock {GC-LSTM:} graph convolution embedded {LSTM} for dynamic link
  prediction.
\newblock {\em CoRR}, abs/1812.04206, 2018.

\bibitem{trivediKnowEvolveDeepTemporal2017}
Rakshit Trivedi, Hanjun Dai, Yichen Wang, and Le~Song.
\newblock Know-evolve: Deep temporal reasoning for dynamic knowledge graphs.
\newblock In Doina Precup and Yee~Whye Teh, editors, {\em Proceedings of the
  34th International Conference on Machine Learning, {ICML} 2017, Sydney, NSW,
  Australia, 6-11 August 2017}, volume~70 of {\em Proceedings of Machine
  Learning Research}, pages 3462--3471. {PMLR}, 2017.

\bibitem{wu2020temp}
Jiapeng Wu, Meng Cao, Jackie Chi~Kit Cheung, and William~L Hamilton.
\newblock {TeMP}: {Temporal} message passing for temporal knowledge graph
  completion.
\newblock {\em arXiv preprint arXiv:2010.03526}, 2020.

\bibitem{liPredictingPathFailure2019}
Jia Li, Zhichao Han, Hong Cheng, Jiao Su, Pengyun Wang, Jianfeng Zhang, and
  Lujia Pan.
\newblock Predicting {{Path Failure In Time}}-{{Evolving Graphs}}.
\newblock In {\em Proceedings of the 25th {{ACM SIGKDD International
  Conference}} on {{Knowledge Discovery}} \& {{Data Mining}}}, {{KDD}} '19,
  pages 1279--1289, {New York, NY, USA}, 2019. {ACM}.

\bibitem{zhou2020heterogeneous}
Fan Zhou, Xovee Xu, Ce~Li, Goce Trajcevski, Ting Zhong, and Kunpeng Zhang.
\newblock A heterogeneous dynamical graph neural networks approach to quantify
  scientific impact.
\newblock {\em arXiv preprint arXiv:2003.12042}, 2020.

\bibitem{wang2020generic}
Yanbang Wang, Pan Li, Chongyang Bai, VS~Subrahmanian, and Jure Leskovec.
\newblock Generic representation learning for dynamic social interaction.
\newblock In {\em The 26th {ACM} {SIGKDD} international conference on knowledge
  discovery \& data mining, {MLG} workshop}, 2020.

\bibitem{zhangDeepLearningGraphs2018}
Ziwei Zhang, Peng Cui, and Wenwu Zhu.
\newblock Deep {{Learning}} on {{Graphs}}: {{A Survey}}.
\newblock {\em arXiv:1812.04202 [cs, stat]}, December 2018.

\bibitem{wuComprehensiveSurveyGraph2019}
Z.~{Wu}, S.~{Pan}, F.~{Chen}, G.~{Long}, C.~{Zhang}, and P.~S. {Yu}.
\newblock A comprehensive survey on graph neural networks.
\newblock {\em IEEE Transactions on Neural Networks and Learning Systems},
  pages 1--21, 2020.

\bibitem{hamiltonRepresentationLearningGraphs2017}
William~L Hamilton, Rex Ying, and Jure Leskovec.
\newblock Representation learning on graphs: {{Methods}} and applications.
\newblock {\em arXiv preprint arXiv:1709.05584}, 2017.

\bibitem{goyalGraphEmbeddingTechniques2018}
Palash Goyal and Emilio Ferrara.
\newblock Graph {{Embedding Techniques}}, {{Applications}}, and
  {{Performance}}: {{A Survey}}.
\newblock {\em Knowledge-Based Systems}, 151:78--94, July 2018.

\bibitem{rossettiCommunityDiscoveryDynamic2018}
Giulio Rossetti and R{\'e}my Cazabet.
\newblock Community {{Discovery}} in {{Dynamic Networks}}: A {{Survey}}.
\newblock {\em ACM Computing Surveys}, 51(2):1--37, February 2018.

\bibitem{kimReviewDynamicNetwork2017}
Bomin Kim, Kevin~H Lee, Lingzhou Xue, and Xiaoyue Niu.
\newblock A review of dynamic network models with latent variables.
\newblock {\em Statistics surveys}, 12:105—135, 2018.

\bibitem{zhangSurveyStreamingAlgorithms2010}
Jian Zhang.
\newblock A survey on streaming algorithms for massive graphs.
\newblock In {\em Managing and {{Mining Graph Data}}}, pages 393--420.
  {Springer}, 2010.

\bibitem{fritzTempusVolatHora2019}
Cornelius Fritz, Michael Lebacher, and G{\"o}ran Kauermann.
\newblock Tempus volat, hora fugit: A survey of tie-oriented dynamic network
  models in discrete and continuous time.
\newblock {\em Statistica Neerlandica}, 2019.

\bibitem{mcgregorGraphStreamAlgorithms2014}
Andrew McGregor.
\newblock Graph stream algorithms: A survey.
\newblock {\em ACM SIGMOD Record}, 43(1):9--20, 2014.

\bibitem{aouayModelingDynamicsSocial2014}
Saoussen Aouay, Salma Jamoussi, Fa{\"i}ez Gargouri, and Ajith Abraham.
\newblock Modeling dynamics of social networks: {{A}} survey.
\newblock In {\em {{CASoN}}}, pages 49--54. {IEEE}, 2014.

\bibitem{rossettiSocialNetworkDynamics2015}
Giulio Rossetti.
\newblock Social {{Network Dynamics}}.
\newblock 2015.

\bibitem{datarMaintainingStreamStatistics2002}
Mayur Datar, Aristides Gionis, Piotr Indyk, and Rajeev Motwani.
\newblock Maintaining {{Stream Statistics}} over {{Sliding Windows}}.
\newblock {\em SIAM Journal on Computing}, 31(6):1794--1813, January 2002.

\bibitem{moriniRevealingEvolutionsDynamical2017}
Matteo Morini, Patrick Flandrin, Eric Fleury, Tommaso Venturini, and Pablo
  Jensen.
\newblock Revealing evolutions in dynamical networks.
\newblock July 2017.

\bibitem{boccalettiStructureDynamicsMultilayer2014}
S.~Boccaletti, G.~Bianconi, R.~Criado, C.~I. {del Genio},
  J.~{G{\'o}mez-Garde{\~n}es}, M.~Romance, I.~{Sendi{\~n}a-Nadal}, Z.~Wang, and
  M.~Zanin.
\newblock The structure and dynamics of multilayer networks.
\newblock {\em Physics Reports}, 544(1):1--122, November 2014.

\bibitem{dunlavyTemporalLinkPrediction2011}
Daniel~M Dunlavy, Tamara~G Kolda, and Evrim Acar.
\newblock Temporal link prediction using matrix and tensor factorizations.
\newblock {\em ACM Transactions on Knowledge Discovery from Data (TKDD)},
  5(2):10, 2011.

\bibitem{barabasiEmergenceScalingRandom1999}
Albert-L{\'a}szl{\'o} Barab{\'a}si and R{\'e}ka Albert.
\newblock Emergence of scaling in random networks.
\newblock {\em science}, 286(5439):509--512, 1999.

\bibitem{leskovecGraphsTimeDensification2005}
Jure Leskovec, Jon Kleinberg, and Christos Faloutsos.
\newblock Graphs over time: Densification laws, shrinking diameters and
  possible explanations.
\newblock In {\em Proceedings of the Eleventh {{ACM SIGKDD}} International
  Conference on {{Knowledge}} Discovery in Data Mining}, pages 177--187. {ACM},
  2005.

\bibitem{youGraphRNNDeepGenerative2018}
Jiaxuan You, Rex Ying, Xiang Ren, William~L. Hamilton, and Jure Leskovec.
\newblock {{GraphRNN}}: {{A Deep Generative Model}} for {{Graphs}}.
\newblock {\em CoRR}, abs/1802.08773, 2018.

\bibitem{newmanNetworks2018}
Mark Newman.
\newblock {\em Networks}.
\newblock {Oxford university press}, 2018.

\bibitem{barabasiEvolutionSocialNetwork2002}
Albert-Laszlo Barab{\^a}si, Hawoong Jeong, Zoltan N{\'e}da, Erzsebet Ravasz,
  Andras Schubert, and Tamas Vicsek.
\newblock Evolution of the social network of scientific collaborations.
\newblock {\em Physica A: Statistical mechanics and its applications},
  311(3-4):590--614, 2002.

\bibitem{xu2015stochastic}
Kevin Xu.
\newblock Stochastic block transition models for dynamic networks.
\newblock In {\em Artificial Intelligence and Statistics}, pages 1079--1087,
  2015.

\bibitem{chenELSTMDDeepLearning2019}
Jinyin Chen, Jian Zhang, Xuanheng Xu, Chengbo Fu, Dan Zhang, Qingpeng Zhang,
  and Qi~Xuan.
\newblock E-{{LSTM}}-{{D}}: {{A Deep Learning Framework}} for {{Dynamic Network
  Link Prediction}}.
\newblock {\em arXiv preprint arXiv:1902.08329}, 2019.

\bibitem{trivedi2018dyrep}
Rakshit Trivedi, Mehrdad Farajtabar, Prasenjeet Biswal, and Hongyuan Zha.
\newblock {{DyRep}}: Learning representations over dynamic graphs.
\newblock In {\em International Conference on Learning Representations}, 2019.

\bibitem{xuDynamicStochasticBlockmodels2014}
Kevin~S. Xu and Alfred~O. Hero~III.
\newblock Dynamic stochastic blockmodels for time-evolving social networks.
\newblock {\em IEEE Journal of Selected Topics in Signal Processing},
  8(4):552--562, August 2014.

\bibitem{ashrafSimulationAugmentationSocial2019}
Akanda Wahid-Ul Ashraf, Marcin Budka, and Katarzyna Musial.
\newblock Simulation and {{Augmentation}} of {{Social Networks}} for {{Building
  Deep Learning Models}}.
\newblock May 2019.

\bibitem{perra2012activity}
Nicola Perra, Bruno Gonçalves, Romualdo Pastor-Satorras, and Alessandro
  Vespignani.
\newblock Activity driven modeling of time varying networks.
\newblock {\em Scientific reports}, 2(1):1--7, 2012.
\newblock Publisher: Nature Publishing Group.

\bibitem{junuthulaEvaluatingLinkPrediction2016}
Ruthwik~R Junuthula, Kevin~S Xu, and Vijay~K Devabhaktuni.
\newblock Evaluating link prediction accuracy in dynamic networks with added
  and removed edges.
\newblock In {\em 2016 {{IEEE}} International Conferences on Big Data and Cloud
  Computing ({{BDCloud}}), Social Computing and Networking ({{SocialCom}}),
  Sustainable Computing and Communications
  ({{SustainCom}})({{BDCloud}}-{{SocialCom}}-{{SustainCom}})}, pages 377--384.
  {IEEE}, 2016.

\bibitem{snijdersIntroductionStochasticActorbased2010}
Tom A.~B. Snijders, Gerhard~G. {van de Bunt}, and Christian E.~G. Steglich.
\newblock Introduction to stochastic actor-based models for network dynamics.
\newblock {\em Social Networks}, 32(1):44--60, January 2010.

\bibitem{divakaran2019temporal}
Aswathy Divakaran and Anuraj Mohan.
\newblock Temporal link prediction: A survey.
\newblock {\em New Generation Computing}, pages 1--46, 2019.

\bibitem{liDeepLearningApproach2014}
Xiaoyi Li, Nan Du, Hui Li, Kang Li, Jing Gao, and Aidong Zhang.
\newblock A deep learning approach to link prediction in dynamic networks.
\newblock In {\em Proceedings of the 2014 {{SIAM International Conference}} on
  {{Data Mining}}}, pages 289--297. {SIAM}, 2014.

\bibitem{liu2013deep}
Feng Liu, Bingquan Liu, Chengjie Sun, Ming Liu, and Xiaolong Wang.
\newblock Deep learning approaches for link prediction in social network
  services.
\newblock In {\em International conference on neural information processing},
  pages 425--432, 2013.
\newblock tex.organization: Springer.

\bibitem{Butts2017}
Carter~T. Butts and Christopher~Steven Marcum.
\newblock A relational event approach to modeling behavioral dynamics.
\newblock In Andrew Pilny and Marshall~Scott Poole, editors, {\em Group
  processes: {Data}-driven computational approaches}, pages 51--92. Springer
  International Publishing, Cham, 2017.

\bibitem{hannekeDiscreteTemporalModels2010}
Steve Hanneke, Wenjie Fu, and Eric~P. Xing.
\newblock Discrete temporal models of social networks.
\newblock {\em Electronic Journal of Statistics}, 4:585--605, 2010.

\bibitem{blockChangeWeCan2018}
Per Block, Johan Koskinen, James Hollway, Christian Steglich, and Christoph
  Stadtfeld.
\newblock Change we can believe in: {{Comparing}} longitudinal network models
  on consistency, interpretability and predictive power.
\newblock {\em Social Networks}, 52:180--191, January 2018.

\bibitem{goldenbergSurveyStatisticalNetwork2010}
Anna Goldenberg, Alice~X. Zheng, Stephen~E. Fienberg, and Edoardo~M. Airoldi.
\newblock A {{Survey}} of {{Statistical Network Models}}.
\newblock {\em Foundations and Trends\textregistered{} in Machine Learning},
  2(2):129--233, February 2010.

\bibitem{kearnesMolecularGraphConvolutions2016}
Steven Kearnes, Kevin McCloskey, Marc Berndl, Vijay Pande, and Patrick Riley.
\newblock Molecular graph convolutions: Moving beyond fingerprints.
\newblock {\em Journal of computer-aided molecular design}, 30(8):595--608,
  2016.

\bibitem{duvenaudConvolutionalNetworksGraphs2015}
David~K Duvenaud, Dougal Maclaurin, Jorge Iparraguirre, Rafael Bombarell,
  Timothy Hirzel, Al{\'a}n {Aspuru-Guzik}, and Ryan~P Adams.
\newblock Convolutional networks on graphs for learning molecular fingerprints.
\newblock In {\em Advances in Neural Information Processing Systems}, pages
  2224--2232, 2015.

\bibitem{ying2018graph}
Rex Ying, Ruining He, Kaifeng Chen, Pong Eksombatchai, William~L Hamilton, and
  Jure Leskovec.
\newblock Graph convolutional neural networks for web-scale recommender
  systems.
\newblock In {\em Proceedings of the 24th {{ACM SIGKDD}} International
  Conference on Knowledge Discovery \& Data Mining}, pages 974--983. {ACM},
  2018.

\bibitem{monti2017geometric}
Federico Monti, Michael Bronstein, and Xavier Bresson.
\newblock Geometric matrix completion with recurrent multi-graph neural
  networks.
\newblock In {\em Advances in Neural Information Processing Systems}, pages
  3697--3707, 2017.

\bibitem{qiuDeepInfSocialInfluence2018}
Jiezhong Qiu, Jian Tang, Hao Ma, Yuxiao Dong, Kuansan Wang, and Jie Tang.
\newblock {{DeepInf}}: {{Social Influence Prediction}} with {{Deep Learning}}.
\newblock {\em Proceedings of the 24th ACM SIGKDD International Conference on
  Knowledge Discovery \& Data Mining - KDD '18}, pages 2110--2119, 2018.

\bibitem{liu2019characterizing}
Yozen Liu, Xiaolin Shi, Lucas Pierce, and Xiang Ren.
\newblock Characterizing and forecasting user engagement with in-app action
  graph: A case study of snapchat.
\newblock In {\em Proceedings of the 25th ACM SIGKDD International Conference
  on Knowledge Discovery \& Data Mining}, pages 2023--2031, 2019.

\bibitem{hamiltonInductiveRepresentationLearning2017}
Will Hamilton, Zhitao Ying, and Jure Leskovec.
\newblock Inductive representation learning on large graphs.
\newblock In {\em Advances in {{Neural Information Processing Systems}}}, pages
  1024--1034, 2017.

\bibitem{fawazDeepLearningTime2019}
Hassan~Ismail Fawaz, Germain Forestier, Jonathan Weber, Lhassane Idoumghar, and
  Pierre-Alain Muller.
\newblock Deep learning for time series classification: A review.
\newblock {\em Data Mining and Knowledge Discovery}, 33(4):917--963, July 2019.

\bibitem{seoStructuredSequenceModeling2018}
Youngjoo Seo, Micha{\"e}l Defferrard, Pierre Vandergheynst, and Xavier Bresson.
\newblock Structured {{Sequence Modeling}} with {{Graph Convolutional Recurrent
  Networks}}.
\newblock In Long Cheng, Andrew Chi~Sing Leung, and Seiichi Ozawa, editors,
  {\em Neural {{Information Processing}}}, Lecture {{Notes}} in {{Computer
  Science}}, pages 362--373. {Springer International Publishing}, 2018.

\bibitem{sarkar2005dynamic}
Purnamrita Sarkar and Andrew Moore.
\newblock Dynamic social network analysis using latent space models.
\newblock In Y.~Weiss, B.~Schölkopf, and J.~Platt, editors, {\em Advances in
  neural information processing systems}, volume~18. MIT Press, 2006.

\bibitem{yang2011detecting}
Tianbao Yang, Yun Chi, Shenghuo Zhu, Yihong Gong, and Rong Jin.
\newblock Detecting communities and their evolutions in dynamic social
  networks—a {Bayesian} approach.
\newblock {\em Machine learning}, 82(2):157--189, 2011.
\newblock Publisher: Springer.

\bibitem{scarselli2008graph}
Franco Scarselli, Marco Gori, Ah~Chung Tsoi, Markus Hagenbuchner, and Gabriele
  Monfardini.
\newblock The graph neural network model.
\newblock {\em IEEE transactions on neural networks}, 20(1):61--80, 2008.
\newblock Publisher: IEEE.

\bibitem{perozziDeepWalkOnlineLearning2014}
Bryan Perozzi, Rami {Al-Rfou}, and Steven Skiena.
\newblock {{DeepWalk}}: {{Online Learning}} of {{Social Representations}}.
\newblock {\em CoRR}, abs/1403.6652, 2014.

\bibitem{defferrardConvolutionalNeuralNetworks2016}
Micha{\"e}l Defferrard, Xavier Bresson, and Pierre Vandergheynst.
\newblock Convolutional {{Neural Networks}} on {{Graphs}} with {{Fast Localized
  Spectral Filtering}}.
\newblock In D.~D. Lee, M.~Sugiyama, U.~V. Luxburg, I.~Guyon, and R.~Garnett,
  editors, {\em Advances in {{Neural Information Processing Systems}} 29},
  pages 3844--3852. {Curran Associates, Inc.}, 2016.

\bibitem{kipfSemisupervisedClassificationGraph2016}
Thomas~N. Kipf and Max Welling.
\newblock Semi-supervised classification with graph convolutional networks.
\newblock In {\em 5th International Conference on Learning Representations,
  {ICLR} 2017, Toulon, France, April 24-26, 2017, Conference Track
  Proceedings}. OpenReview.net, 2017.

\bibitem{manessiDynamicGraphConvolutional2020}
Franco Manessi, Alessandro Rozza, and Mario Manzo.
\newblock Dynamic graph convolutional networks.
\newblock {\em Pattern Recognition}, 97:107000, January 2020.

\bibitem{kumar2019predicting}
Srijan Kumar, Xikun Zhang, and Jure Leskovec.
\newblock Predicting dynamic embedding trajectory in temporal interaction
  networks.
\newblock In {\em Proceedings of the 25th ACM SIGKDD International Conference
  on Knowledge Discovery \& Data Mining}, pages 1269--1278, 2019.

\bibitem{maStreamingGraphNeural2018}
Yao Ma, Ziyi Guo, Zhaochun Ren, Eric Zhao, Jiliang Tang, and Dawei Yin.
\newblock Streaming {{Graph Neural Networks}}.
\newblock {\em arXiv:1810.10627 [cs, stat]}, October 2018.

\bibitem{xuGenerativeGraphConvolutional2019}
Da~Xu, Chuanwei Ruan, Kamiya Motwani, Evren Korpeoglu, Sushant Kumar, and
  Kannan Achan.
\newblock Generative {{Graph Convolutional Network}} for {{Growing Graphs}}.
\newblock {\em ICASSP 2019 - 2019 IEEE International Conference on Acoustics,
  Speech and Signal Processing (ICASSP)}, pages 3167--3171, May 2019.

\bibitem{kipfVariationalGraphAutoencoders2016}
Thomas~N. Kipf and Max Welling.
\newblock Variational graph auto-encoders.
\newblock {\em CoRR}, abs/1611.07308, 2016.

\bibitem{qu_continuous-time_2020}
Liang Qu, Huaisheng Zhu, Qiqi Duan, and Yuhui Shi.
\newblock Continuous-time link prediction via temporal dependent graph neural
  network.
\newblock In {\em Proceedings of the web conference 2020}, {WWW} '20, pages
  3026--3032, New York, NY, USA, 2020. Association for Computing Machinery.
\newblock Number of pages: 7 Place: Taipei, Taiwan.

\bibitem{gersLearningPreciseTiming}
Felix~A Gers, Nicol~N Schraudolph, and Jurgen Schmidhuber.
\newblock Learning {{Precise Timing}} with {{LSTM Recurrent Networks}}.
\newblock page~29.

\bibitem{narayanLearningGraphDynamics2018}
Apurva Narayan and Peter~H. O'N~Roe.
\newblock Learning {{Graph Dynamics}} using {{Deep Neural Networks}}.
\newblock {\em IFAC-PapersOnLine}, 51(2):433--438, January 2018.

\bibitem{niepertLearningConvolutionalNeural2016}
Mathias Niepert, Mohamed Ahmed, and Konstantin Kutzkov.
\newblock Learning convolutional neural networks for graphs.
\newblock In Maria{-}Florina Balcan and Kilian~Q. Weinberger, editors, {\em
  Proceedings of the 33nd International Conference on Machine Learning, {ICML}
  2016, New York City, NY, USA, June 19-24, 2016}, volume~48 of {\em {JMLR}
  Workshop and Conference Proceedings}, pages 2014--2023. JMLR.org, 2016.

\bibitem{taheriLearningRepresentEvolution2019}
Aynaz Taheri, Kevin Gimpel, and Tanya {Berger-Wolf}.
\newblock Learning to {{Represent}} the {{Evolution}} of {{Dynamic Graphs}}
  with {{Recurrent Models}}.
\newblock In {\em Companion {{Proceedings}} of {{The}} 2019 {{World Wide Web
  Conference}}}, {{WWW}} '19, pages 301--307, {New York, NY, USA}, 2019. {ACM}.

\bibitem{liGatedGraphSequence2017}
Yujia Li, Daniel Tarlow, Marc Brockschmidt, and Richard~S. Zemel.
\newblock Gated graph sequence neural networks.
\newblock In Yoshua Bengio and Yann LeCun, editors, {\em 4th International
  Conference on Learning Representations, {ICLR} 2016, San Juan, Puerto Rico,
  May 2-4, 2016, Conference Track Proceedings}, 2016.

\bibitem{hochreiterLongShortTermMemory1997}
Sepp Hochreiter and J{\"u}rgen Schmidhuber.
\newblock Long {{Short}}-{{Term Memory}}.
\newblock {\em Neural Comput.}, 9(8):1735--1780, November 1997.

\bibitem{velickovicGraphAttentionNetworks2017}
Petar Velickovic, Guillem Cucurull, Arantxa Casanova, Adriana Romero, Pietro
  Li{\`{o}}, and Yoshua Bengio.
\newblock Graph attention networks.
\newblock In {\em 6th International Conference on Learning Representations,
  {ICLR} 2018, Vancouver, BC, Canada, April 30 - May 3, 2018, Conference Track
  Proceedings}. OpenReview.net, 2018.

\bibitem{vaswaniAttentionAllYou2017}
Ashish Vaswani, Noam Shazeer, Niki Parmar, Jakob Uszkoreit, Llion Jones,
  Aidan~N Gomez, {\L}ukasz Kaiser, and Illia Polosukhin.
\newblock Attention is {{All}} you {{Need}}.
\newblock In I.~Guyon, U.~V. Luxburg, S.~Bengio, H.~Wallach, R.~Fergus,
  S.~Vishwanathan, and R.~Garnett, editors, {\em Advances in {{Neural
  Information Processing Systems}} 30}, pages 5998--6008. {Curran Associates,
  Inc.}, 2017.

\bibitem{wang2020tedic}
Yanbang Wang, Pan Li, Chongyang Bai, and Jure Leskovec.
\newblock {TEDIC}: {Neural} modeling of behavioral patterns in dynamic social
  interaction networks.
\newblock 2020.

\bibitem{wavenet}
Aäron van~den Oord, Sander Dieleman, Heiga Zen, Karen Simonyan, Oriol Vinyals,
  Alexander Graves, Nal Kalchbrenner, Andrew Senior, and Koray Kavukcuoglu.
\newblock {WaveNet}: {A} generative model for raw audio.
\newblock In {\em Arxiv}, 2016.

\bibitem{zhang_link_2018}
Muhan Zhang and Yixin Chen.
\newblock Link prediction based on graph neural networks.
\newblock In {\em Advances in neural information processing systems}, pages
  5165--5175, 2018.

\bibitem{cai2020structural}
Lei Cai, Zhengzhang Chen, Chen Luo, Jiaping Gui, Jingchao Ni, Ding Li, and
  Haifeng Chen.
\newblock Structural temporal graph neural networks for anomaly detection in
  dynamic graphs.
\newblock {\em arXiv preprint arXiv:2005.07427}, 2020.

\bibitem{shiConvolutionalLSTMNetwork2015}
Xingjian Shi, Zhourong Chen, Hao Wang, Dit{-}Yan Yeung, Wai{-}Kin Wong, and
  Wang{-}chun Woo.
\newblock Convolutional {LSTM} network: {A} machine learning approach for
  precipitation nowcasting.
\newblock In Corinna Cortes, Neil~D. Lawrence, Daniel~D. Lee, Masashi Sugiyama,
  and Roman Garnett, editors, {\em Advances in Neural Information Processing
  Systems 28: Annual Conference on Neural Information Processing Systems 2015,
  December 7-12, 2015, Montreal, Quebec, Canada}, pages 802--810, 2015.

\bibitem{jin2019recurrent}
Woojeong Jin, He~Jiang, Meng Qu, Tong Chen, Changlin Zhang, Pedro Szekely, and
  Xiang Ren.
\newblock Recurrent event network: Global structure inference over temporal
  knowledge graph.
\newblock {\em arXiv: 1904.05530}, 2019.

\bibitem{bonner2019temporal}
Stephen Bonner, Amir Atapour-Abarghouei, Philip~T Jackson, John Brennan, Ibad
  Kureshi, Georgios Theodoropoulos, Andrew~Stephen McGough, and Boguslaw Obara.
\newblock Temporal neighbourhood aggregation: {Predicting} future links in
  temporal graphs via recurrent variational graph convolutions.
\newblock In {\em 2019 {IEEE} international conference on big data (big data)},
  pages 5336--5345, 2019.
\newblock tex.organization: IEEE.

\bibitem{schlichtkrullModelingRelationalData2018}
Michael Schlichtkrull, Thomas~N. Kipf, Peter Bloem, Rianne {van den Berg}, Ivan
  Titov, and Max Welling.
\newblock Modeling {{Relational Data}} with {{Graph Convolutional Networks}}.
\newblock In Aldo Gangemi, Roberto Navigli, Maria-Esther Vidal, Pascal Hitzler,
  Rapha{\"e}l Troncy, Laura Hollink, Anna Tordai, and Mehwish Alam, editors,
  {\em The {{Semantic Web}}}, Lecture {{Notes}} in {{Computer Science}}, pages
  593--607. {Springer International Publishing}, 2018.

\bibitem{goyal2018dynamicgem}
Palash Goyal, Sujit~Rokka Chhetri, Ninareh Mehrabi, Emilio Ferrara, and
  Arquimedes Canedo.
\newblock {{DynamicGEM}}: {{A}} library for dynamic graph embedding methods.
\newblock {\em arXiv preprint arXiv:1811.10734}, 2018.

\bibitem{wangStructuralDeepNetwork2016}
Daixin Wang, Peng Cui, and Wenwu Zhu.
\newblock Structural {{Deep Network Embedding}}.
\newblock In {\em Proceedings of the {{22Nd ACM SIGKDD International
  Conference}} on {{Knowledge Discovery}} and {{Data Mining}}}, {{KDD}} '16,
  pages 1225--1234, {New York, NY, USA}, 2016. {ACM}.

\bibitem{chen2015net2net}
Tianqi Chen, Ian~J. Goodfellow, and Jonathon Shlens.
\newblock Net2net: Accelerating learning via knowledge transfer.
\newblock In Yoshua Bengio and Yann LeCun, editors, {\em 4th International
  Conference on Learning Representations, {ICLR} 2016, San Juan, Puerto Rico,
  May 2-4, 2016, Conference Track Proceedings}, 2016.

\bibitem{goyalDyngraph2vecCapturingNetwork2019}
Palash Goyal, Sujit~Rokka Chhetri, and Arquimedes Canedo.
\newblock Dyngraph2vec: {{Capturing Network Dynamics}} using {{Dynamic Graph
  Representation Learning}}.
\newblock {\em Knowledge-Based Systems}, page S0950705119302916, July 2019.

\bibitem{hajiramezanaliVariationalGraphRecurrent2019}
Ehsan Hajiramezanali, Arman Hasanzadeh, Krishna~R. Narayanan, Nick Duffield,
  Mingyuan Zhou, and Xiaoning Qian.
\newblock Variational graph recurrent neural networks.
\newblock In Hanna~M. Wallach, Hugo Larochelle, Alina Beygelzimer, Florence
  d'Alch{\'{e}}{-}Buc, Emily~B. Fox, and Roman Garnett, editors, {\em Advances
  in Neural Information Processing Systems 32: Annual Conference on Neural
  Information Processing Systems 2019, NeurIPS 2019, 8-14 December 2019,
  Vancouver, BC, Canada}, pages 10700--10710, 2019.

\bibitem{yinSemiImplicitVariationalInference2018}
Mingzhang Yin and Mingyuan Zhou.
\newblock Semi-{{Implicit Variational Inference}}.
\newblock In {\em International {{Conference}} on {{Machine Learning}}}, pages
  5660--5669, July 2018.

\bibitem{goodfellowGenerativeAdversarialNets2014}
Ian Goodfellow, Jean {Pouget-Abadie}, Mehdi Mirza, Bing Xu, David
  {Warde-Farley}, Sherjil Ozair, Aaron Courville, and Yoshua Bengio.
\newblock Generative {{Adversarial Nets}}.
\newblock In Z.~Ghahramani, M.~Welling, C.~Cortes, N.~D. Lawrence, and K.~Q.
  Weinberger, editors, {\em Advances in {{Neural Information Processing
  Systems}} 27}, pages 2672--2680. {Curran Associates, Inc.}, 2014.

\bibitem{wang2019generative}
Zhengwei Wang, Qi~She, and Tomas~E Ward.
\newblock Generative adversarial networks: A survey and taxonomy.
\newblock {\em arXiv preprint arXiv:1906.01529}, 2019.

\bibitem{lei2019gcn}
Kai Lei, Meng Qin, Bo~Bai, Gong Zhang, and Min Yang.
\newblock Gcn-gan: A non-linear temporal link prediction model for weighted
  dynamic networks.
\newblock In {\em IEEE INFOCOM 2019-IEEE Conference on Computer
  Communications}, pages 388--396. IEEE, 2019.

\bibitem{xiong2019dyngraphgan}
Yun Xiong, Yao Zhang, Hanjie Fu, Wei Wang, Yangyong Zhu, and S~Yu Philip.
\newblock Dyngraphgan: Dynamic graph embedding via generative adversarial
  networks.
\newblock In {\em International Conference on Database Systems for Advanced
  Applications}, pages 536--552. Springer, 2019.

\bibitem{baytasPatientSubtypingTimeAware2017}
Inci~M. Baytas, Cao Xiao, Xi~Zhang, Fei Wang, Anil~K. Jain, and Jiayu Zhou.
\newblock Patient {{Subtyping}} via {{Time}}-{{Aware LSTM Networks}}.
\newblock In {\em Proceedings of the 23rd {{ACM SIGKDD International
  Conference}} on {{Knowledge Discovery}} and {{Data Mining}} - {{KDD}} '17},
  pages 65--74, {Halifax, NS, Canada}, 2017. {ACM Press}.

\bibitem{knyazev2019learning}
Boris Knyazev, Carolyn Augusta, and Graham~W Taylor.
\newblock Learning temporal attention in dynamic graphs with bilinear
  interactions.
\newblock {\em arXiv preprint arXiv:1909.10367}, 2019.

\bibitem{kipf2018neural}
Thomas~N. Kipf, Ethan Fetaya, Kuan{-}Chieh Wang, Max Welling, and Richard~S.
  Zemel.
\newblock Neural relational inference for interacting systems.
\newblock In Jennifer~G. Dy and Andreas Krause, editors, {\em Proceedings of
  the 35th International Conference on Machine Learning, {ICML} 2018,
  Stockholmsm{\"{a}}ssan, Stockholm, Sweden, July 10-15, 2018}, volume~80 of
  {\em Proceedings of Machine Learning Research}, pages 2693--2702. {PMLR},
  2018.

\bibitem{han2020graph}
Zhen Han, Yuyi Wang, Yunpu Ma, Stephan Günnemann, and Volker Tresp.
\newblock The graph hawkes network for reasoning on temporal knowledge graphs.
\newblock {\em arXiv preprint arXiv:2003.13432}, 2020.

\bibitem{mei2017neural}
Hongyuan Mei and Jason~M Eisner.
\newblock The neural hawkes process: {A} neurally self-modulating multivariate
  point process.
\newblock In {\em Advances in neural information processing systems}, pages
  6754--6764, 2017.

\bibitem{kazemi2019time2vec}
Seyed~Mehran Kazemi, Rishab Goel, Sepehr Eghbali, Janahan Ramanan, Jaspreet
  Sahota, Sanjay Thakur, Stella Wu, Cathal Smyth, Pascal Poupart, and Marcus
  Brubaker.
\newblock Time2vec: {Learning} a vector representation of time.
\newblock {\em arXiv preprint arXiv:1907.05321}, 2019.

\bibitem{xu_self-attention_2019}
Da~Xu, Chuanwei Ruan, Evren Korpeoglu, Sushant Kumar, and Kannan Achan.
\newblock Self-attention with functional time representation learning.
\newblock In H.~Wallach, H.~Larochelle, A.~Beygelzimer, F.~dAlché Buc, E.~Fox,
  and R.~Garnett, editors, {\em Advances in neural information processing
  systems}, volume~32. Curran Associates, Inc., 2019.

\bibitem{wu2020evonet}
Changmin Wu, Giannis Nikolentzos, and Michalis Vazirgiannis.
\newblock {EvoNet}: {A} neural network for predicting the evolution of dynamic
  graphs.
\newblock {\em arXiv preprint arXiv:2003.00842}, 2020.

\bibitem{yangEvaluatingLinkPrediction2015}
Yang Yang, Ryan~N. Lichtenwalter, and Nitesh~V. Chawla.
\newblock Evaluating link prediction methods.
\newblock {\em Knowl. Inf. Syst.}, 45(3):751--782, 2015.

\bibitem{aalen2008survival}
Odd Aalen, Ornulf Borgan, and Hakon Gjessing.
\newblock {\em Survival and Event History Analysis: A Process Point of View}.
\newblock {Springer Science \& Business Media}, 2008.

\bibitem{liDeepDynamicNetwork2018}
Taisong Li, Jiawei Zhang, S~Yu Philip, Yan Zhang, and Yonghong Yan.
\newblock Deep dynamic network embedding for link prediction.
\newblock {\em IEEE Access}, 6:29219--29230, 2018.

\bibitem{luLinkPredictionComplex2011}
Linyuan L{\"u} and Tao Zhou.
\newblock Link prediction in complex networks: {{A}} survey.
\newblock {\em Physica A}, 390(6):11501170, 2011.

\bibitem{martinezSurveyLinkPrediction2016}
V{\'i}ctor Mart{\'i}nez, Fernando Berzal, and Juan-Carlos Cubero.
\newblock A {{Survey}} of {{Link Prediction}} in {{Complex Networks}}.
\newblock {\em ACM Comput. Surv.}, 49(4):69:1--69:33, December 2016.

\bibitem{DeepGCNsCanGCNs}
Guohao Li, Matthias M{\"{u}}ller, Ali~K. Thabet, and Bernard Ghanem.
\newblock Deepgcns: Can gcns go as deep as cnns?
\newblock In {\em 2019 {IEEE/CVF} International Conference on Computer Vision,
  {ICCV} 2019, Seoul, Korea (South), October 27 - November 2, 2019}, pages
  9266--9275. {IEEE}, 2019.

\end{thebibliography}

\begin{IEEEbiography}[{\includegraphics[width=1in,height=1.25in,clip,keepaspectratio]{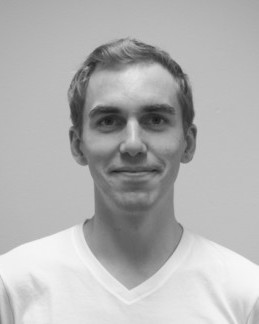}}]{Joakim Skarding} received the M.Sc degree in computer science from the Norwegian University of Science and Technology (NTNU) in 2015.

He worked as a software developer for TomTom in Berlin and since 2018 he has been pursuing his PhD at the University of Technology Sydney. His research interest include modelling and prediction of dynamic complex networks.
 
\end{IEEEbiography}

\begin{IEEEbiography}[{\includegraphics[width=1in,height=1.25in,clip,keepaspectratio]{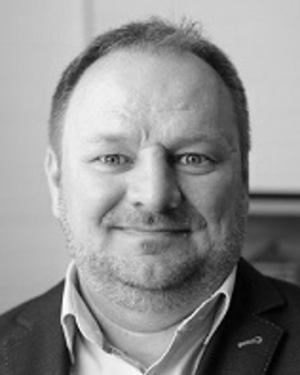}}]{Bogdan Gabrys} received the M.Sc. degree in electronics and telecommunication from Silesian Technical University, Gliwice, Poland, in 1994, and the
Ph.D. degree in computer science from Nottingham
Trent University, Nottingham, U.K., in 1998.

Over the last 25 years, he has been working at
various universities and research and development
departments of commercial institutions. He is currently a Professor of Data Science and the Director of the Advanced Analytics Institute at the University of
Technology Sydney, Sydney, Australia. His research
activities have concentrated on the areas of data science, complex adaptive
systems, computational intelligence, machine learning, predictive analytics,
and their diverse applications. He has published over 180 research papers,
chaired conferences, workshops, and special sessions, and been on program
committees of a large number of international conferences with the data science, computational intelligence, machine learning, and data mining themes.
He is also a Senior Member of the Institute of Electrical and Electronics
Engineers (IEEE), a Member of IEEE Computational Intelligence Society and
a Fellow of the Higher Education Academy (HEA) in the UK. He is frequently
invited to give keynote and plenary talks at international conferences and
lectures at internationally leading research centres and commercial research
labs. More details can be found at: http://bogdan-gabrys.com
\end{IEEEbiography}

\begin{IEEEbiography}[{\includegraphics[width=1in,height=1.25in,clip,keepaspectratio]{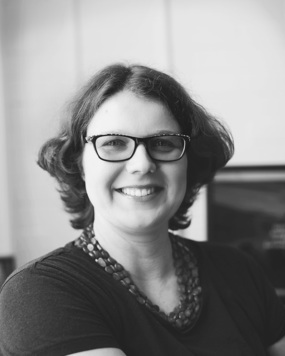}}]{KATARZYNA MUSIAL} received the M.Sc. degree in computer science from the Wrocław University of Science and Technology (WrUST), Poland, the M.Sc. degree in software engineering from the Blekinge Institute of Technology, Sweden, in 2006, and the Ph.D. from WrUST, in November 2009. 

In November 2009, she was appointed as a Senior Visiting Research Fellow with Bournemouth University (BU), where she has been a Lecturer in informatics, since 2010. In November 2011, she joined Kings as a Lecturer in computer science. In September 2015, she returned to Bournemouth University as a Principal Academic in Computing, where she was a member of the Data Science Institute. In September 2017, she joined as an Associate Professor in network science with the School of Software, University of Technology Sydney, where she is currently a member of the Advanced Analytics Institute. Her research interests include complex networked systems, analysis of their dynamics and its evolution, adaptive and predictive modeling of their structure and characteristics, as well as the adaptation mechanisms that exist within such systems are in the center of her research interests. 
\end{IEEEbiography}

\begin{table*}[hpbt]
\caption{\label{tab:notation} A summary of notation used in this work. }
\centering
\begin{tabular}{ll}
Notation & Description\\
\hline
\(\odot\) & Element wise product\\
\(G\) & Static graph\\
\(G^t\) & Static graph at time \(t\)\\
\(DG\) & Discrete dynamic graph\\
\(CG\) & Continuous dynamic graph\\
\(V\) & The set of nodes in a graph\\
\(E\) & The set of edges in a graph\\
\(v\) & A node \(v \in V\)\\
\(e\) & An edge \(e \in E\)\\
\(t\) & Time step / event time\\
\(\bar{t}\) & Time step just before time t\\
\({<t}\) & All time steps up until time t\\
\(\Delta\) & Duration\\
\(n\) & Number of nodes\\
\(d\) & Dimensions of a node feature vector\\
\(l\) & Dimensions of a GNN produced hidden feature vector\\
\(k\) & Dimensions of a RNN/self-attention produced hidden feature vector\\
\(X^t\) & Feature matrix at time \(t\)\\
\(A^t\) & Adjacency matrix at time \(t\)\\
\(\hat{A}\) & Predicted adjacency matrix\\
\(z_u^t\) & GNN produced hidden feature vector of node \(u\) at time \(t\)\\
\(h_u^t\) & RNN/self-attention produced hidden feature vector of node \(u\) at time \(t\)\\
\(Z^t\) & GNN produced hidden feature matrix at time \(t\)\\
\(H^t\) & RNN/self-attention produced hidden feature matrix at time \(t\)\\
\(\lambda\) & Conditional intensity function\\
\(\sigma\) & The sigmoid function\\
\(W, U, w, b, \omega, \psi, \varphi\) & Learnable model parameters\\
\end{tabular}
\end{table*}

\begin{table*}
\caption{\label{tab:abbreviation} A summary of abbreviations used in this work. Some model names in the text look like abbreviations but are in fact simply the name of the model (or the authors do not explicitly state what the abbreviation stand for). These include: PATCHY-SAN, DyGGNN, RgGNN, StrGNN, EvolveGCN, JODIE, GC-LSTM, GCN-GAN, DynGraphGAN and DyREP.}
\centering
\begin{tabular}{ll}
\makecell{Abbreviation} & Description\\
\hline
GNN & Graph neural network\\
DGNN & Dynamic graph neural network\\
RNN & Recurrent neural network\\
LSTM & Long-term short term memory\\
GRU & Gated recurrent unit\\
GAN & Generative adversarial network\\
CNN & Convolutional neural network\\
TPP & Temporal point process\\
\hline
RGM & Random graph model\\
ERGM & Exponential random graph model\\
TERGM & Temporal exponential random graph model\\
SAOM & Stochastic actor oriented model\\
REM & Relational event model\\
\hline
GCN & Graph Convolutional Network \cite{kipfSemisupervisedClassificationGraph2016}\\
GAT & Graph Attention Network \cite{velickovicGraphAttentionNetworks2017}\\
GGNN & Gated Graph Neural Network \cite{liGatedGraphSequence2017}\\
R-GCN & Relational Graph Convolutional Network \cite{schlichtkrullModelingRelationalData2018}\\
convLSTM & Convolutional LSTM \cite{shiConvolutionalLSTMNetwork2015}\\
GraphRNN & Graph recurrent neural network \cite{youGraphRNNDeepGenerative2018}\\
G-GCN & Generative graph convolutional network \cite{xuGenerativeGraphConvolutional2019}\\
VGAE & Variational graph autoencoder \cite{kipfVariationalGraphAutoencoders2016}\\
\hline
GCRN-M1 & Graph convolutional recurrent network model 1 \cite{seoStructuredSequenceModeling2018}\\
GCRN-M2 & Graph convolutional recurrent network model 2 \cite{seoStructuredSequenceModeling2018}\\
WD-GCN & Waterfall dynamic graph convolutional network \cite{manessiDynamicGraphConvolutional2020}\\
CD-GCN & Concatenated dynamic graph convolutional network \cite{manessiDynamicGraphConvolutional2020}\\
DySAT & Dynamic Self-Attention Network \cite{sankarDynamicGraphRepresentation2018}\\
TNDCN & Temporal network-diffusion convolution networks\cite{wang2020generic}\cite{wang2020tedic}\\
HDGNN & Heterogeneous Dynamical Graph Neural Network\cite{zhou2020heterogeneous}\\
TeMP & Temporal Message Passing \cite{wu2020temp}\\
LRGCN & Long Short-Term Memory R-GCN \cite{liPredictingPathFailure2019} \\
RE-Net & Recurrent Event Network \cite{jin2019recurrent}\\
TNA & Temporal Neighbourhood Aggregation\cite{bonner2019temporal}\\
TDGNN & Temporal Dependent Graph Neural Network \cite{qu_continuous-time_2020}\\
DynGEM & Dynamic graph embedding model \cite{goyal2018dynamicgem}\\
E-LSTM-D & Encode-LSTM-Decode \cite{chenELSTMDDeepLearning2019}\\
VGRNN & Variational graph recurrent neural network \cite{yinSemiImplicitVariationalInference2018}\\
SI-VGRNN & Semi-implicit VGRNN \cite{yinSemiImplicitVariationalInference2018}\\
SGNN & Streaming graph neural network \cite{maStreamingGraphNeural2018}\\
LDG & Latent dynamic graph \cite{knyazev2019learning}\\
GHN & Graph Hawkes network \cite{han2020graph}\\
TGAT & Temporal graph attention \cite{xu2020inductive}\\
TGN & Temporal Graph Networks \cite{rossi2020temporal}\\
\hline

\end{tabular}
\end{table*}

\EOD

\end{document}